\documentclass[twocolumn]{aastex631}
\let\tablenum\relax
\usepackage{booktabs} 
\usepackage{siunitx}    
\usepackage{multirow}   
\usepackage{amsmath}    
\usepackage{graphicx}    
\usepackage{makecell}

\shorttitle{Mock Catalog of lensed GWs}
\shortauthors{Sun et al.}

\begin{document}

\title{Mock Catalogs of Strongly Lensed Gravitational Waves via a Halo Model Approach with Space-borne Detectors}

\correspondingauthor{Kai Liao}
\email{liaokai@whu.edu.cn}

\author{Mingqi Sun}
\affiliation{School of Physics and Technology, Wuhan University, Wuhan 430072, China}

\author{Kai Liao*}
\affiliation{School of Physics and Technology, Wuhan University, Wuhan 430072, China}

\author{Youkai Li}
\affiliation{School of Physics and Technology, Wuhan University, Wuhan 430072, China}

\author{Tonghua Liu}
\affiliation{School of Physics and Optoelectronic Engineering, Yangtze University, Jingzhou 434023, China}

\author{Hengyu  Wu}
\affiliation{School of Physics and Optoelectronic Engineering, Yangtze University, Jingzhou 434023, China}

\author{Shaoqi Hou}
\affiliation{School of Physics and Technology, Wuhan University, Wuhan 430072, China}

\author{Tao Yang}
\affiliation{School of Physics and Technology, Wuhan University, Wuhan 430072, China}

\author{Xilong Fan}
\affiliation{School of Physics and Technology, Wuhan University, Wuhan 430072, China}

\author{Marek Biesiada}
\affiliation{National Centre for Nuclear Research, Pasteura 7, PL-02-093 Warsaw, Poland;}

\begin{abstract}
Future space-borne gravitational-wave (GW) detectors, such as LISA and DECIGO, are expected to detect a large number of GW events, a fraction of which may be strongly lensed by intervening galaxies or galaxy clusters. In this work, we develop a comprehensive framework to simulate strongly lensed GWs in the context of space-borne detectors. Based on realistic astrophysical models for both the source population and the lens distribution, we construct mock catalogs of lensed GW events, referred to as \textbf{GW-LMC-Space}. Our results show that, for a four-year LISA observation, the expected number of lensed events ranges from $0$ to $131$, depending on the adopted formation model of massive black hole binaries (MBHBs). The corresponding lensing probability for MBHBs can reach up to $\sim 0.3\%$. For DECIGO, we find that the number of lensed events in a one-year observation is expected to lie in the range of $0$--$44$, with a lensing probability of $\sim 0.15\%$ for stellar-mass binary black holes (BBHs), binary neutron stars (BNSs), and neutron star--black hole binaries (NSBHs). We further show that the overlap of lensed signals is a common feature in space-borne detectors, which can significantly affect both the signal-to-noise ratio (SNR) estimation and event identification. These results highlight the importance of accounting for signal overlap in the analysis of strongly lensed GW events in future space-borne GW observations.
\end{abstract}


\section{Introduction}
\label{sec:intro}

As a direct prediction of general relativity, gravitational waves (GWs) have become a powerful probe of extreme astrophysical events since their first detection in 2015 \citep{2016PhRvL.116f1102A}, corresponding to the binary black hole merger GW150914. To date, 218 GW events have been reported by the LIGO--Virgo--KAGRA collaboration \citep{2025arXiv250818082T,2025arXiv250904348T}, with many more expected in the coming years. These observations mark the advent of gravitational-wave astronomy, providing unprecedented opportunities to test fundamental physics, investigate the nature of compact objects, and probe the large-scale properties of the Universe \citep{2020GReGr..52...81B,2019PhRvD.100h4002H,2026PhRvL.136j1003C}.

Another key prediction of general relativity is gravitational lensing, which has been extensively observed in the electromagnetic domain. Massive objects acting as lenses can deflect light from distant sources, producing multiple images with magnifications and time delays. This phenomenon provides a powerful probe of dark matter distributions, cosmological parameters, and the nature of gravity \citep{2025PhRvD.112b3002S,2025PhRvD.112b3546L,2020MNRAS.494.1956M,2019ApJ...880...50Y,2020MNRAS.497L..56Y,2025arXiv250509507Y}.
Gravitational waves can likewise be lensed as they propagate through the gravitational field of massive objects. A key difference from electromagnetic waves is that GWs have much longer wavelengths, such that wave-optics effects may become important in certain regimes \citep{2003ApJ...595.1039T}. In this work, we focus on the geometric-optics regime.

For current ground-based detectors, previous studies have found no statistically significant evidence for strongly lensed GW events in the LIGO--Virgo network during the O4 run \citep{2025arXiv251216347T}. However, with the next generation of detectors, such as Cosmic Explorer (CE) \citep{2021arXiv210909882E} and the Einstein Telescope (ET) \citep{2025arXiv250312263A}, strongly lensed GW events are expected to be detected in the near future. There are many applications with these signals, for example, measuring the Hubble constant \cite{2017NatCo...8.1148L}. It has been estimated that on the order of several hundred lensed events (e.g., $\sim 600$) could be observed over the next decade \citep{li2026mockcatalogsstronglylensed}. 

Space-based GW detectors, including Taiji \citep{2020IJMPA..3550075R}, TianQin \citep{2016CQGra..33c5010L}, DECIGO \citep{2011PhRvD..83d4011Y}, and LISA \citep{2017arXiv170200786A}, have not yet been launched. Nevertheless, their distinct frequency bands and long observation times will provide a complementary window for studying gravitational lensing of GWs. Early studies suggest that a small but non-negligible number of strongly lensed events may be detectable during a multi-year LISA mission \citep{2010PhRvL.105y1101S}. More optimistic assessments of $\sim$60 lensed GW signals per year were made for the DECIGO in \cite{Piorkowska2021}.

In this work, we construct mock catalogs of strongly lensed GWs for the LISA and DECIGO detectors, whose sensitive frequency bands are approximately $10^{-4}$--$10^{-1}$ Hz and $0.1$--$10$ Hz, respectively. The dominant sources in these bands include massive black hole binaries (MBHBs), stellar-mass binary black holes (BBHs), binary neutron stars (BNSs), and neutron star--black hole binaries (NSBHs). We adopt a halo-based lensing framework to model the lensing effects on these sources and compute the corresponding signal-to-noise ratios (SNRs) of lensed events in order to assess their detectability.

Compared to previous studies \citep{2025PhRvD.112l3512G}, a key feature of our work is the explicit treatment of GW signal duration. For space-based detectors, the signal duration can be comparable to or even exceed typical lensing time delays, leading to potential overlap between lensed signals. This effect has important implications for SNR estimation and event identification, and is therefore incorporated into our analysis.

This paper is organized as follows. In Section~\ref{sec:lensing_model}, we summarize the main components of the lens model based on the SL-Hammocks framework. In Section~\ref{sec:lisa}, we describe the MBHB source model and present the mock catalog of strongly lensed GWs detectable by LISA. In Section~\ref{sec:decigo}, we perform a similar analysis for DECIGO, focusing on stellar-mass BBHs, BNSs, and NSBHs. Finally, we present our conclusions in Section~\ref{sec:conclusion}. Throughout this paper, we adopt a flat $\Lambda$CDM cosmology consistent with the latest Planck results \citep{2020A&A...641A...6P}. All the catalogs have been published on the Github\footnote{\url{https://github.com/LensedGW/GW-LMC-Space}} and Zenodo\footnote{\url{https://doi.org/10.5281/zenodo.20540877}}\citep{lensedgw_2026_20540878}.

\section{Lens Model}\label{sec:lensing_model}

Our lensing framework is based on the new-generation code \textbf{Strong Lensing Halo Model--based Mock Catalogs (SL-Hammocks)} \citep{2025OJAp....8E...8A}, which has been publicly released on GitHub\footnote{\url{https://github.com/LSSTDESC/SL-Hammocks}}. In contrast to the traditional \textbf{OM10} model \citep{OM10}, which is restricted to galaxy-scale lenses, the halo-based formalism implemented in SL-Hammocks provides a self-consistent description of lens systems across a broad mass range, from sub-galaxy-scale to cluster-scale halos. In this section, we summarize the main ingredients of the lens model relevant to our analysis.

\subsection{Host Dark Matter Halos and Central Galaxies}

The population of host dark matter halos is generated using the halo mass function defined in terms of the virial mass from \citet{2008ApJ...688..709T}. All calculations are performed using the \textbf{COLOSSUS} package \citep{2018ApJS..239...35D}. 

The mass distribution of host halos is modeled by the Navarro--Frenk--White (NFW; \citealt{1996ApJ...462..563N}) density profile,
\begin{equation}
    \rho_{\mathrm{hh}}(r) = \frac{\rho_{\mathrm{s,hh}}}{(r/r_{\mathrm{s,hh}})\left(1+r/r_{\mathrm{s,hh}}\right)^2}\,,
\end{equation}
where $\rho_{\mathrm{s,hh}}$ and $r_{\mathrm{s,hh}}$ denote the characteristic density and scale radius, respectively. These parameters are determined by the host halo mass $M_{\mathrm{hh}}$ and concentration parameter $c_{\mathrm{hh}}$ through
\begin{equation}
    \rho_{\mathrm{s,hh}} = \frac{M_{\mathrm{hh}}}{4\pi r_{\mathrm{s,hh}}^{3}\, m_{\mathrm{nfw}}(c_{\mathrm{hh}})}\,,
\end{equation}
with
\begin{equation}
    m_{\mathrm{nfw}}(c) = \ln(1+c) - \frac{c}{1+c}\,.
\end{equation}

The concentration parameter $c_{\mathrm{hh}}$ is assumed to follow a lognormal distribution with dispersion $\sigma_{\ln c_{\mathrm{hh}}}=0.33$ \citep{Maccio2007}, while the parameter $\mu$ denotes the mean value of $\ln c_{\mathrm{hh}}$, which comes from the mass-concentration relation presented by \cite{2019ApJ...871..168D},
\begin{equation}
p(c_{\mathrm{hh}}) = \frac{1}{c_{\mathrm{hh}}\sqrt{2\pi}\sigma_{\ln c_{\mathrm{hh}}}}
\exp\left[-\frac{\left(\ln c_{\mathrm{hh}} - \mu\right)^2}
{2\sigma_{\ln c_{\mathrm{hh}}}^2}\right]\,.
\end{equation}

Ellipticity is incorporated by replacing the spherical radius $r$ in the NFW profile with the elliptical radius
\begin{equation}
    v \equiv \sqrt{\frac{\tilde{x}^2}{1-e_{\mathrm{hh}}}+(1-e_{\mathrm{hh}})\tilde{y}^2}\,,
\end{equation}
where $(\tilde{x},\tilde{y})$ are rotated coordinates defined by
\begin{equation}
    \begin{aligned}
    \tilde{x}=x\cos{\phi_{\mathrm{hh}}} + y \sin{\phi_{\mathrm{hh}}}\,,\\
    \tilde{y}=y\cos{\phi_{\mathrm{hh}}}-x\sin{\phi_{\mathrm{hh}}}\,.
    \end{aligned}
\end{equation}
The ellipticity $e_{\mathrm{hh}}$ follows a truncated normal distribution, while the position angle $\phi_{\mathrm{hh}}$ is assumed to be uniformly distributed. Detailed prescriptions can be found in \citet{2020MNRAS.496.2591O, 2025OJAp....8E...8A}.

Each host halo is assumed to contain a central galaxy modeled by an elliptical Hernquist profile,
\begin{equation}
    \rho_{c*}(v_{c*}) =
    \frac{f_{c*}M_{\mathrm{hh}}}
    {2\pi (v_{c*}/r_{b,\mathrm{cen}})(v_{c*}+r_{b,\mathrm{cen}})^3}\,,
\end{equation}
where $f_{c*}$ denotes the stellar mass fraction. The scale radius is given by $r_{b,\mathrm{cen}}=0.551\,r_{\mathrm{e,cen}}$, with $r_{\mathrm{e,cen}}$ determined following \citet{2024ApJ...960...53V}. The stellar mass fraction is obtained from the stellar-to-halo mass relation (SHMR) calibrated in \citet{2013ApJ...762L..31B, 2019MNRAS.488.3143B}. Since the SHMR depends on the assumed initial mass function (IMF), we adopt the Salpeter IMF throughout this work.

A key advantage of this halo-based framework is that no scale-dependent assumptions are imposed on the lens population. The same physical prescription consistently applies to galaxy-scale and cluster-scale systems, enabling a self-consistent and physically motivated treatment of all lensing phenomena considered in this paper, including lensed gravitational waves.

\subsection{Subhalos and External Shear}

The model further incorporates subhalos and their associated satellite galaxies. The subhalo mass function at redshift $z$, accounting for tidal stripping and dynamical friction, is given by \citet{2020ApJ...901...58O},
\begin{equation}
    \frac{dN_{\mathrm{sh}}}{dM_{\mathrm{sh}}}
    = f_{\mathrm{df}}
    \frac{dN_{\mathrm{sh}}}{dM_{\mathrm{f}}}
    \frac{dM_{\mathrm{f}}}{dM_{\mathrm{sh}}}\,,
\end{equation}
where $M_{\mathrm{f}}$ is the subhalo mass at accretion and $f_{\mathrm{df}}$ represents the suppression factor due to dynamical friction. The progenitor mass function is predicted by the extended Press–Schechter formalism,
\begin{equation}
\frac{dN_{\mathrm{sh}}}{dM_{\mathrm{f}}}
= \frac{M_{\mathrm{hh}}}{M_{\mathrm{f}}}
P(M_{\mathrm{f}},z_{\mathrm{f}}\,|M_{\mathrm{hh}},z)\,.
\end{equation}

The radial distribution of subhalos is assumed to approximately trace the mass distribution of the host halo. Positions are generated using inverse transform sampling and subsequently projected onto the lens plane to obtain the two-dimensional spatial distribution.

Although the subhalo mass function incorporates tidal stripping effects (i.e., formally adopting truncated NFW profiles), \citet{2025OJAp....8E...8A} employ untruncated elliptical NFW profiles when computing lensing quantities for numerical stability and simplicity. In this implementation, $M_{\mathrm{f}}$ is treated as the effective subhalo mass. This approximation is justified because strong lensing signals are primarily sensitive to the inner density profile and are only weakly affected by the stripped outer regions.

The concentration–mass relation for subhalos is adopted from \citet{2020MNRAS.492.3662I} and compared with that for host halos from \citet{2015ApJ...799..108D}, with the larger value assigned in each case. Each subhalo is further associated with a satellite galaxy, whose stellar mass follows the satellite stellar mass–halo mass (SMHM) relation given in Appendix~J of \citet{2019MNRAS.488.3143B}.

As for the external shear, $\gamma_{\mathrm{ext}} = \sqrt{\gamma_{\mathrm{ext,1}}^2 + \gamma_{\mathrm{ext,2}}^2}$, the code adopts both $\gamma_{\mathrm{ext,1}}$ and $\gamma_{\mathrm{ext,2}}$ are following a truncated normal distribution with dispersion of $\sigma_{\gamma_{\mathrm{ext}}}$, which is redshift-dependent. Additionally, the external convergence is excluded from the lensing calculation for simplicity.

\subsection{Lensing Calculation}
The lensing code \texttt{SL-Hammocks} evaluates lensing effects using the public code \texttt{glafic} \citep{2021PASP..133g4504O}. Rather than solving the lens equation directly by including the lensing potentials of all deflectors simultaneously, \texttt{SL-Hammocks} decomposes the full lens population into individual deflector systems and solves the lens equation for each system separately. When a source lies within the Einstein radius of a given deflector, the corresponding lensing effect is then computed. To perform the lensing calculation, we further input the source population and its signal-to-noise ratio (SNR) in each redshift bin. This separation between the lens and source populations is possible because the lens modeling implemented in \texttt{SL-Hammocks} is independent of the source properties. The lens parameters adopted in the calculation are summarized in Table~1 of \citet{2025OJAp....8E...8A}.

\section{Detector: LISA}\label{sec:lisa}

In this section, we describe the procedure for constructing mock catalogs of strongly lensed GWs detectable by LISA. We focus on modeling the source population and computing the SNR. We then present the resulting mock catalog and discuss the expected number of lensed GW events detectable during the nominal LISA mission.

\subsection{Source Model}\label{sec:formation_models}

The primary science targets of LISA are massive black hole binaries (MBHBs), which constitute the most promising sources in the millihertz frequency band. Although significant uncertainties remain in the formation and evolutionary pathways of MBHBs, we adopt the latest simulation results from \cite{2012MNRAS.423.2533B,2025PhRvD.112l3512G} to generate the intrinsic MBHB population.

Two main classes of seed formation scenarios are considered. The first is the light-seed model (PopIII), in which black hole seeds of mass $10^{2}$--$10^{3}\,M_{\odot}$ form at redshifts $z \sim 15$--$20$. The second is the heavy-seed model (HS/Q3), in which seed masses are $10^{4}$--$10^{5}\,M_{\odot}$ and form at redshifts $z \sim 8$--$15$. In total, six specific realizations are adopted: \texttt{HSnodnoSN}, \texttt{HSnodSN}, \texttt{HSnodSNhighaccr}, \texttt{PopIIId}, \texttt{Q3d}, and \texttt{Q3nod}. Detailed descriptions of these models can be found in \cite{2016PhRvD..93b4003K,2015ApJ...812...72A} and \cite{2020ApJ...904...16B,2023PhRvD.108j3034B}.

In this naming convention, the labels ``nod'' and ``d'' indicate whether the delay time between the host-galaxy merger and the MBHB coalescence is neglected or included, respectively. The labels ``SN'' and ``noSN'' denote whether supernova feedback is taken into account.

According to \cite{2025PhRvD.112l3512G}, the intrinsic number of MBHB merger events during a four-year LISA mission, together with the subset satisfying $\mathrm{SNR}>8$, are summarized in Table~\ref{c}. 

The detection threshold is applied within the adopted catalog prescription.
A finite mission window can affect the SNR, the observable duration, and which
delayed images contribute to a detected system; the model-dependent sensitivity
to this limitation is quantified in Section~\ref{sec:conclusion}.

\begin{deluxetable}{lcc}
\tablecaption{Intrinsic and detectable event numbers of MBHBs for different formation models in 4-year LISA observation. Detectable  number means the event of $\rho_0 > 8$.}
\label{c}
\tablehead{
\colhead{Model} &
\colhead{Intrinsic number} &
\colhead{Detectable number } 
}
\startdata
HSnodnoSN & 39364 & 37979 \\
HSnodSN & 36790 & 35510 \\
HSnodSNhighaccr &9407&9075 \\
PopIIId &1410&321 \\
Q3d &74&73\\
Q3nod &657&650
\enddata
\end{deluxetable}

\begin{figure*}
    \centering
    \label{Figure: intinsic event snr}
    \includegraphics[width=1.0\linewidth]{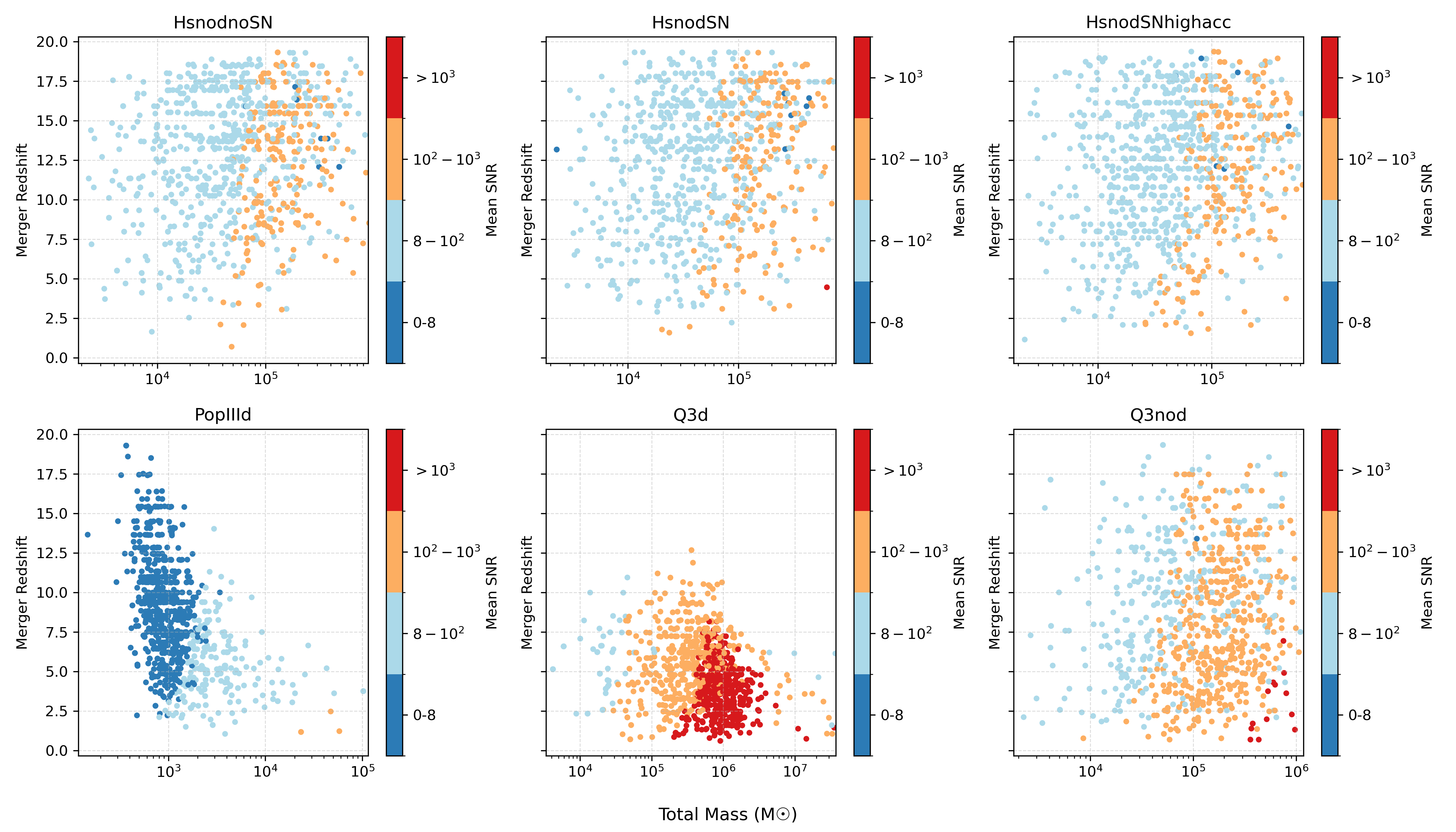}
    \caption{SNR distribution (encoded in color map) of the intrinsic population of MBHBs for different formation models in 4-year LISA observation. The "x" axis is the total mass of the system, while the "y" axis denotes the redshift of the source (merging MBHB). For each model, we randomly selected 1000 events, considered their event rate from the intrinsic population and ploted the SNR distribution.}
\end{figure*}

The SNR is computed using the \texttt{IMRPhenomHM} waveform model in combination with the LISA noise power spectral density. More precisely, we used the SNR calculation tool - \texttt{LISAbeta}, which accounts for the confusion noise coming from unresolved galactic binaries\citep{2021PhRvD.103h3011M}.
For each intrinsic merger event, 100 realizations of the extrinsic parameters (coalescence time, coalescence phase, inclination angle, polarization angle, and sky location) are randomly sampled, yielding a total of 10{,}000 simulated events per model \citep{2025PhRvD.112l3512G}. The resulting SNR distributions are shown in Figure~\ref{Figure: intinsic event snr}. The full SNR catalogs for all six models are publicly available\footnote{\url{https://github.com/juangutimuni/Stronglensing-LISA/tree/main/Catalogs}}.

\subsection{Lensed GW Catalog: LISA Case}
We now apply the lensing framework described in Section~2 to the intrinsic MBHB population to construct a mock catalog of strongly lensed gravitational waves detectable by LISA. The resulting catalog includes both intrinsic and lensed SNRs for each event, together with the corresponding lensing quantities, such as magnification and time delay.

Before presenting the simulation results, it is important to clarify a key feature of the LISA detector. In contrast to ground-based detectors, gravitational-wave signals from MBHBs in the LISA band can persist for months to years. By comparison, the typical time delays predicted by our lensing model are on the order of days to weeks. Consequently, lensed signals cannot always be treated as a set of independent images, as is commonly assumed for ground-based detectors. Instead, whether multiple images appear as independent signals or as a single overlapping signal depends on the comparison between the GW signal duration and the lensing time delay.

To quantify this effect, we estimate the duration of each GW signal. Following \cite{1963PhRv..131..435P}, the leading-order evolution of the gravitational-wave frequency can be written as
\begin{equation}
    \label{frequency_evolution}
\frac{df}{dt}=
\frac{96}{5}\pi^{8/3}
\left(\frac{G\mathcal{M}}{c^3}\right)^{5/3}
f^{11/3}\,,
\end{equation}
where $\mathcal{M}$ denotes the chirp mass of the binary system. Integrating this equation yields the full-band duration of the inspiral signal,
\begin{equation}
    \label{delta_t_obs}
\Delta t_{\mathrm{obs}}=
\frac{5}{256}
\left(\frac{G\mathcal{M}_z}{c^3}\right)^{-5/3}
\pi^{-8/3}
\left(
f_{\mathrm{low}}^{-8/3}
-
f_{\mathrm{ISCO}}^{-8/3}
\right)\,,
\end{equation}
where $\mathcal{M}_z$ is the redshifted chirp mass, $f_{\mathrm{low}}$ is the lower frequency cutoff of LISA (typically $10^{-4}\,\mathrm{Hz}$), and $f_{\mathrm{ISCO}}$ is the innermost stable circular orbit frequency,
\begin{equation}
    \label{ISCO}
f_{\mathrm{ISCO}}=
\frac{c^3}{6^{3/2}\pi G \mathcal{M}_z}\,.
\end{equation}

The duration estimated in Eq.~(\ref{delta_t_obs}) is the full-band inspiral duration, rather
than the duration necessarily observed during the mission. The latter also
depends on the position of the signal within the observing window and cannot
exceed the mission duration. In particular, a full-band duration exceeding
four years does not imply that the corresponding SNR accumulates uniformly
over that interval.

If $f_{\mathrm{low}} > f_{\mathrm{ISCO}}$, the inspiral–merger–ringdown signal does not enter the LISA frequency band. The above expressions describe only the inspiral phase. Since the final coalescence and ringdown stage has a much shorter characteristic timescale compared to the inspiral phase, its contribution to the total signal duration is neglected in this estimate.

\begin{figure}
    \includegraphics[width=1.0\linewidth]{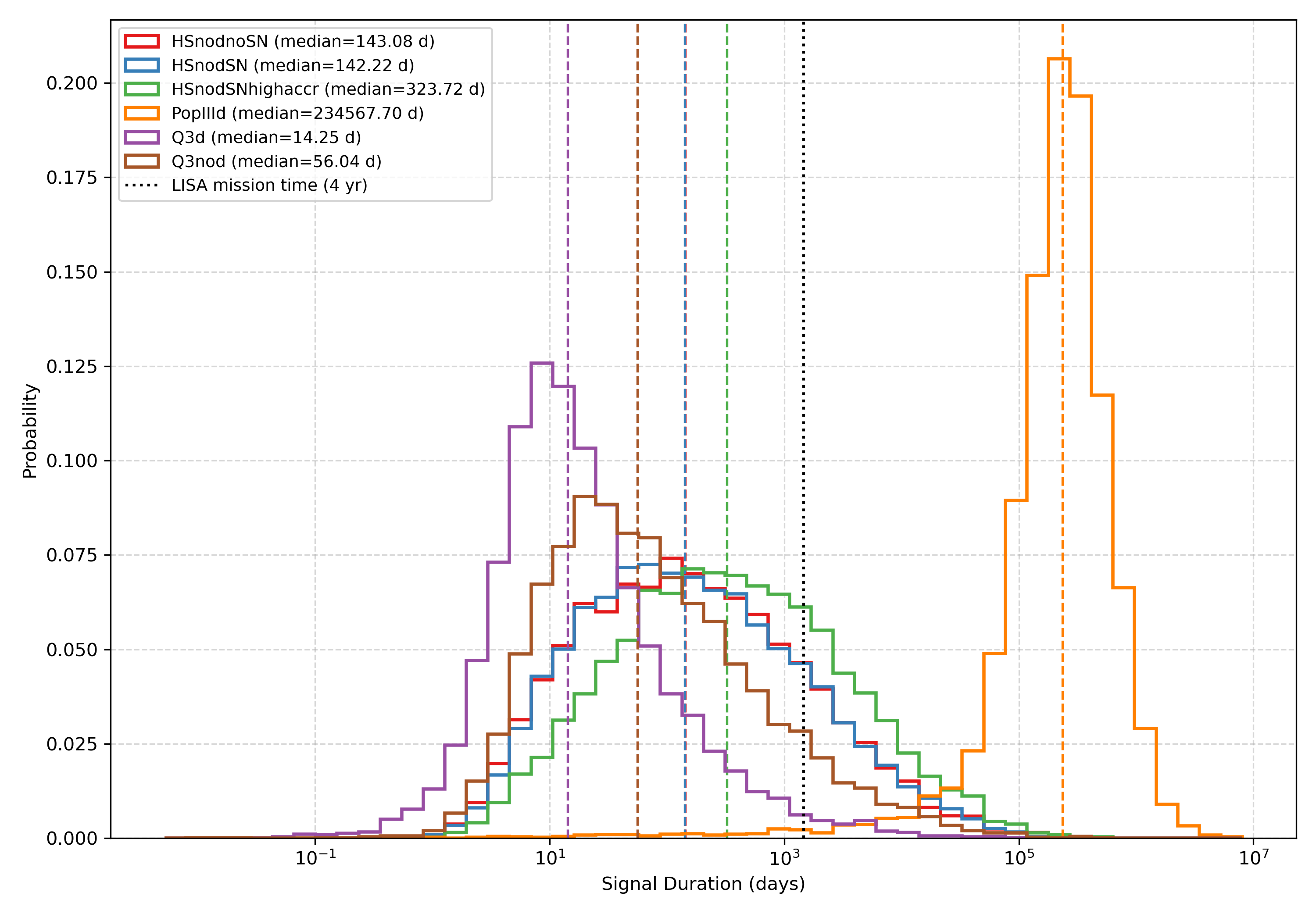}
    \caption{Distribution of the full-band inspiral durations for the 10,000 entries in each MBHB source catalog, calculated using Eqs.~(\ref{frequency_evolution})--(\ref{ISCO}). The vertical black line marks the nominal four-year LISA mission duration. These full-band durations are used in the catalog overlap prescription and are not truncated to an event-dependent observation window; see Section~\ref{sec:conclusion}.}
    \label{Figure:duration_time_distribution}
\end{figure}

Applying these equations to each simulated event yields the distribution of GW signal durations shown in Figure~\ref{Figure:duration_time_distribution}. The six formation models all exhibit a broad distribution of signal durations, indicating that the relationship between signal duration and lensing time delay must be carefully treated when computing the SNR of lensed events.

We therefore adopt the following strategy:

\begin{enumerate}
\item If the lensing time delay is shorter than the GW signal duration, the lensed signals overlap in time and are treated as a single combined signal. The SNR is then computed from the full lensed waveform.
\item If the lensing time delay exceeds the GW signal duration, the images are treated as independent signals, and the SNR of each image is computed separately.
\end{enumerate}

The corresponding optimal SNR expressions are
\begin{equation}
\rho^{L}_{\mathrm{full}}=
\sqrt{\langle h^{L}(f)\,|\,h^{L}(f)\rangle}\,,
\label{snr_fullwaveform}
\end{equation}
\begin{equation}
\rho^{L}_{i}=
\sqrt{\mu_i}\,\rho_0\,,
\end{equation}
where $\rho^{L}_{\mathrm{full}}$ is the SNR of the combined lensed signal, $h^{L}(f)$ is the lensed waveform, $\rho^{L}_{i}$ is the SNR of the $i$-th image, $\mu_i$ denotes the magnification factor, and $\rho_0$ is the SNR of the unlensed signal. The noise-weighted inner product is defined as \citep{2003ApJ...595.1039T}:
\begin{equation}
\langle h^{L}(f)\,|\,h^{L}(f)\rangle
=
4
\int_0^{\infty}
\frac{|h^{L}(f)|^2}{S_n(f)}\,df\,,
\end{equation}
where $S_n(f)$ is the detector noise power spectral density. The lensed waveform can be expressed as
\begin{equation}
h^{L}(f)=
\left(
\sum_j
|\mu_j|^{1/2}
\exp\left[i2\pi f t_{d,j}-i\pi n_j\right]
\right)h(f)\,,
\label{lensed_waveform}
\end{equation}
where $t_{d,j}$ is the time delay of the $j$-th image and $n_j$ denotes the corresponding Morse index. All these quantities are obtained from the lensing simulation described in Section~2.

We then apply the lens model to the intrinsic MBHB population. For clarity, we first compute the number of lensed events without imposing any SNR detection threshold (commonly taken as $\rho = 8$), which allows us to directly evaluate the lensing probability of each formation model. The resulting numbers are summarized in Table~\ref{tab:catalog_lensed_event_number_original}.

We find an overall lensing probability of approximately $0.3\%$, which is slightly smaller than the $\sim 0.5\%$ reported in \cite{2025PhRvD.112l3512G}. This difference arises because our work employs a realistic lensing simulation to generate individual lens systems, whereas the previous study estimated the lensing probability using an optical-depth approximation with simple Singular Isothermal Sphere (SIS) model.

Table~\ref{tab:catalog_lensed_event_number_original} also presents the number of lensed events with different numbers of observable signals, both with and without accounting for signal overlap. We will refer to the scenario neglecting signal overlap as baseline. Although many lensing systems formally produce multiple images, the number of independent signals is significantly reduced when overlap between lensed signals is taken into account. This demonstrates that signal overlap is a crucial factor that must be considered when analyzing strongly lensed GW events in the LISA band.

\begin{deluxetable*}{lcccccc} 
    \tablecaption{Baseline lensed event number of MBHBs without threshold for different formation models in 4-year LISA observation} 
    \label{tab:catalog_lensed_event_number_original} 
    \tablehead{ \colhead{Model} & \colhead{Total number} & \colhead{1-signal} & \colhead{2-signals} & \colhead{3-signals} & \colhead{4-signals} & \colhead{5-signals} } 
    \startdata 
    HSnodnoSN & 131 & 21(98) & 11(28) & 85(4) &3(1) & 11(0) \\ 
    HSnodSN & 106 & 22(80) & 7(20) & 73(6) &0(0) & 4(0) \\ 
    HSnodSNhighaccr & 29 & 5 (25)& 1(3) & 20(1) &1(0) & 2(0) \\ 
    PopIIId & 4 & 0(4) & 1(0) & 3(0) &0(0) & 0(0) \\ 
    Q3d & 0 & 0 & 0 & 0 &0 & 0\\ 
    Q3nod & 2 & 1(2) & 0(0) & 1(0) &0(0) & 0(0) 
    \enddata 
    \tablecomments{The numbers in the parentheses represents the number of lensed events when the
signal overlap is considered.  {The predicted number of intrinsic events during the 4-year LISA observation varies significantly among different formation models. For the models HSnodnoSN, HSnodSN, HSnodSNhighaccr, PopIIId, Q3d, and Q3nod, the predicted numbers are 39364, 36790, 9407, 1410, 74, and 657, respectively, as listed in Table~\ref{c}.}}
\end{deluxetable*}

Finally, we note that the number of lensed events predicted by some models is significantly smaller than that of others. To better understand the lensing behavior of different formation scenarios, we randomly sample $10^5$ sets of intrinsic parameters from each model and apply the lensing simulation to these samples in order to obtain more robust statistical information of the lensed event rates. To account for the stochasticity of the lensing simulation, we repeat this procedure five times for each model, generating five independent catalogs. All catalogs are publicly available on GitHub we have mentioned above. In this work, we use one representative catalog to present our results.

\begin{figure}
    \centering
    \includegraphics[width=1.0\linewidth]{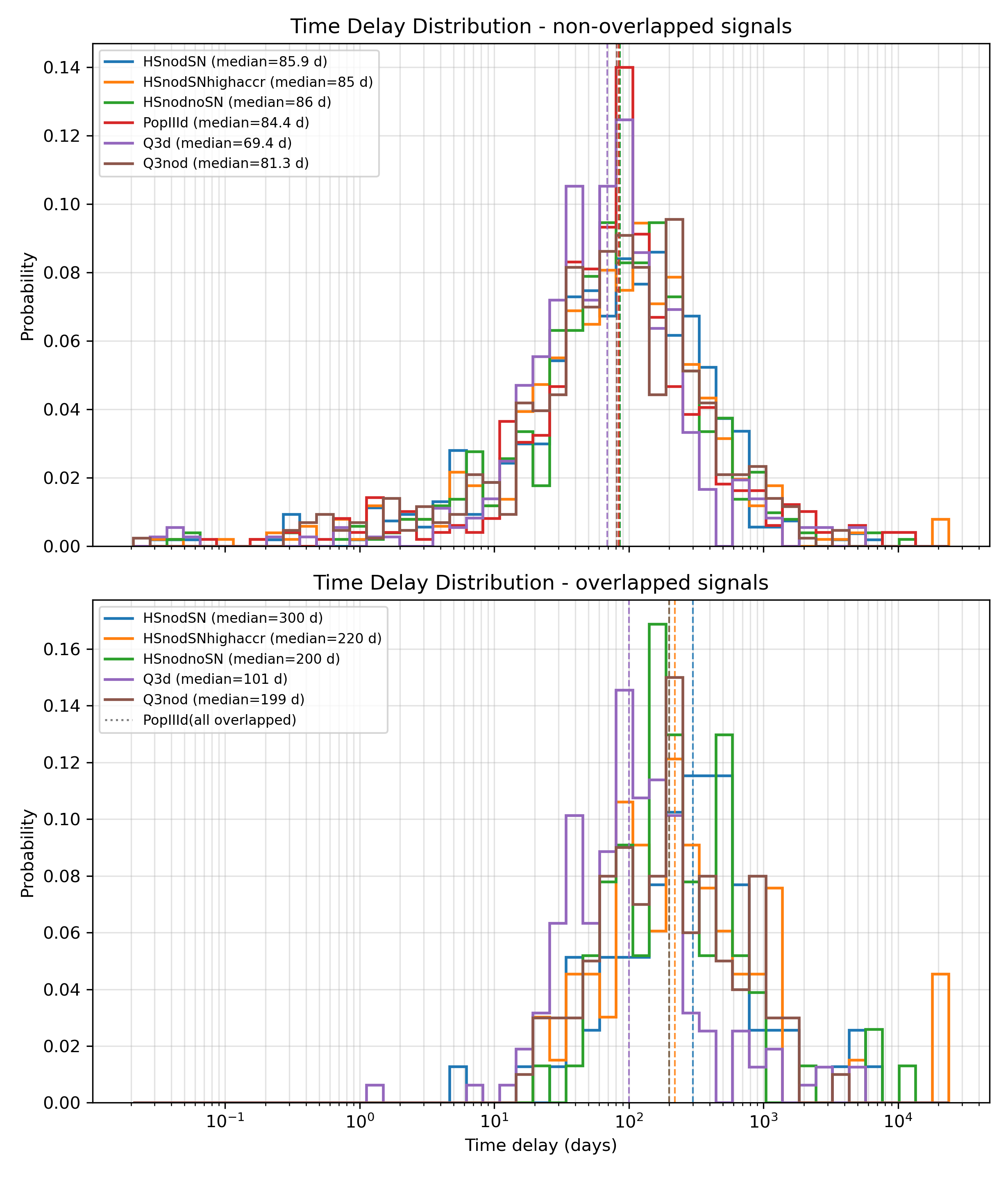}
    \caption{Time delay distribution of the lensed events for different formation models in 100,000 samples of LISA observation. The upper panel shows the time delay distribution without considering the overlapping of lensed signals, while the lower panel shows the distribution when we consider the overlapping.}
    \label{Figure:time_delay_distribution}
\end{figure}

We first present the time-delay distribution of lensed events obtained from $10^5$ samples for each model, both with and without accounting for signal overlap, as shown in Figure~\ref{Figure:time_delay_distribution}. Without considering overlap, the time delays span a broad range, from a few days to several months, with most being shorter than the corresponding signal duration, consistent with the results in Table~\ref{tab:catalog_lensed_event_number_original}. This further highlights the importance of accounting for signal overlap when analyzing lensed GW events in the LISA band.

As shown in the lower panel of Figure~\ref{Figure:time_delay_distribution}, the effective time delays become significantly longer when overlap is taken into account. This is because lensed signals with shorter time delays are more likely to overlap and be identified as a single signal, whereas those with longer time delays are more likely to be distinguished as independent signals. This effect also implies that the observed distribution of time delays for detected lensed events will be biased toward longer values due to signal overlap.

\begin{deluxetable*}{lccccccccc} 
    \tablecaption{Statistical results of detectable lensed events consigering 100,000 samples for different formation models.} 
    \label{tab:catalog_lensed_event_number_final} 
    \tablehead{ \colhead{Model} & \colhead{Detected number} & \colhead{$N_{1}$} & \colhead{$N_{2}$} & \colhead{$N_{3}$} & \colhead{$N_{4}$} & \colhead{$N_{5}$} & \colhead{$N_{sub}$} & \colhead{$N_{r_{\mathrm{max}}>3}$} & \colhead{$N_{r_{\mathrm{max}}>10}$} } 
    \startdata 
    HSnodSN           & 330 & 274 & 39 & 14 & 1 & 2 & 98 & 27 & 1 \\
    HSnodSNhighaccr   & 322 & 264 & 51 & 6  & 1 & 0 & 82 & 25 & 0 \\
    HSnodnoSN         & 319 & 256 & 50 & 13 & 0 & 0 & 65 & 27 & 0 \\
    PopIIId           & 169 & 169 & 0  & 0  & 0 & 0 & 49 & 23 & 0 \\
    Q3d               & 214 & 111 & 52 & 46 & 4 & 1 & 61 & 12 & 0 \\
    Q3nod             & 253 & 177 & 56 & 18 & 0 & 2 & 84 & 21 & 0 \\
    \enddata 
    \tablecomments{$N_{1}$, $N_{2}$, $N_{3}$, $N_{4}$, and $N_{5}$ represent the number of events with 1, 2, 3, 4, and 5 detectable signals, respectively. $N_{sub}$ denotes the number of events lensed by subhalos, while $N_{r_{\mathrm{max}}>3}$ and $N_{r_{\mathrm{max}}>10}$ represent the number of events with max magnification factors greater than 3 and 10, respectively.}
\end{deluxetable*}

In Table~\ref{tab:catalog_lensed_event_number_final}, we summarize the statistical results for the detected lensed events in each formation model when signal overlap is taken into account. We define a ``detected lensing system'' as one in which the maximum SNR among all overlapping signals exceeds the detection threshold (set to 8), thereby providing an optimistic estimate and maximizing the number of multiple-signal systems.

This criterion counts a system if at least one of its signal groups exceeds
the threshold; it does not establish that lensing is identifiable or that all
geometrical images are detectable within the mission. The finite-window
sensitivity of these counts and the limitations of the overlap classification
are discussed in Section~\ref{sec:conclusion}.

The total number of detected lensed events ranges from 169 to 330 across different models, with the majority producing a single detectable signal. However, a non-negligible fraction of events yield multiple detectable signals due to lensing, with some systems producing up to five detectable images. The number of events lensed by subhalos is also significant, indicating that substructure lensing can play an important role in shaping the observed properties of lensed GWs.

\begin{figure*}
    \centering
    \includegraphics[width=1.0\linewidth]{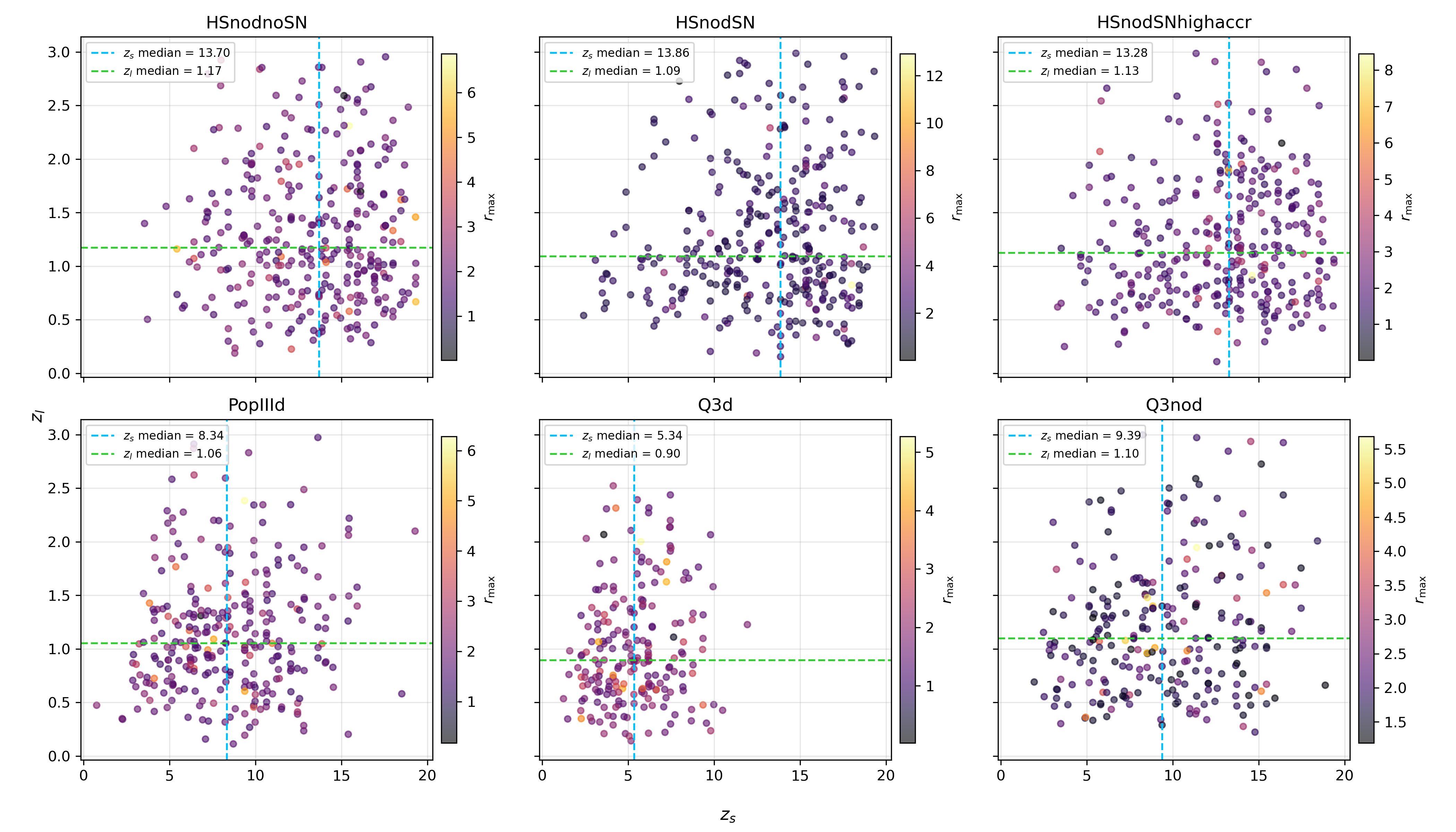}
    \caption{We present the distribution of lens redshift $z_l$ (redshift of the host halo) and source redshift $z_s$ for $10^5$ samples, together with a scatter plot of the maximum SNR ratio $r_{\mathrm{max}}$ (defined as the ratio of the SNR of the overlapped signal to that of the unlensed signal) for different formation models in a four-year LISA observation. The color bar indicates the value of $r_{\mathrm{max}}$, while the blue and green dashed lines denote the median source redshift and lens redshift, respectively.}
    \label{Figure: snr_distribution}
\end{figure*}

Furthermore, because signal overlap is taken into account, the conventional magnification factor $\mu$ is no longer directly applicable. Instead, we define an effective magnification as the ratio of the SNR of the overlapped signal to that of the unlensed signal, denoted as $r$. We find that a significant fraction of lensed events have maximum magnification factors greater than 3 ($r_{\mathrm{max}} > 3$), while a smaller subset exhibits extreme magnifications ($r_{\mathrm{max}} > 10$). The distribution of the maximum magnification $r_{\mathrm{max}}$ is shown in Figure~\ref{Figure: snr_distribution}.

\begin{figure}
    \centering
    \includegraphics[width=1.0\linewidth]{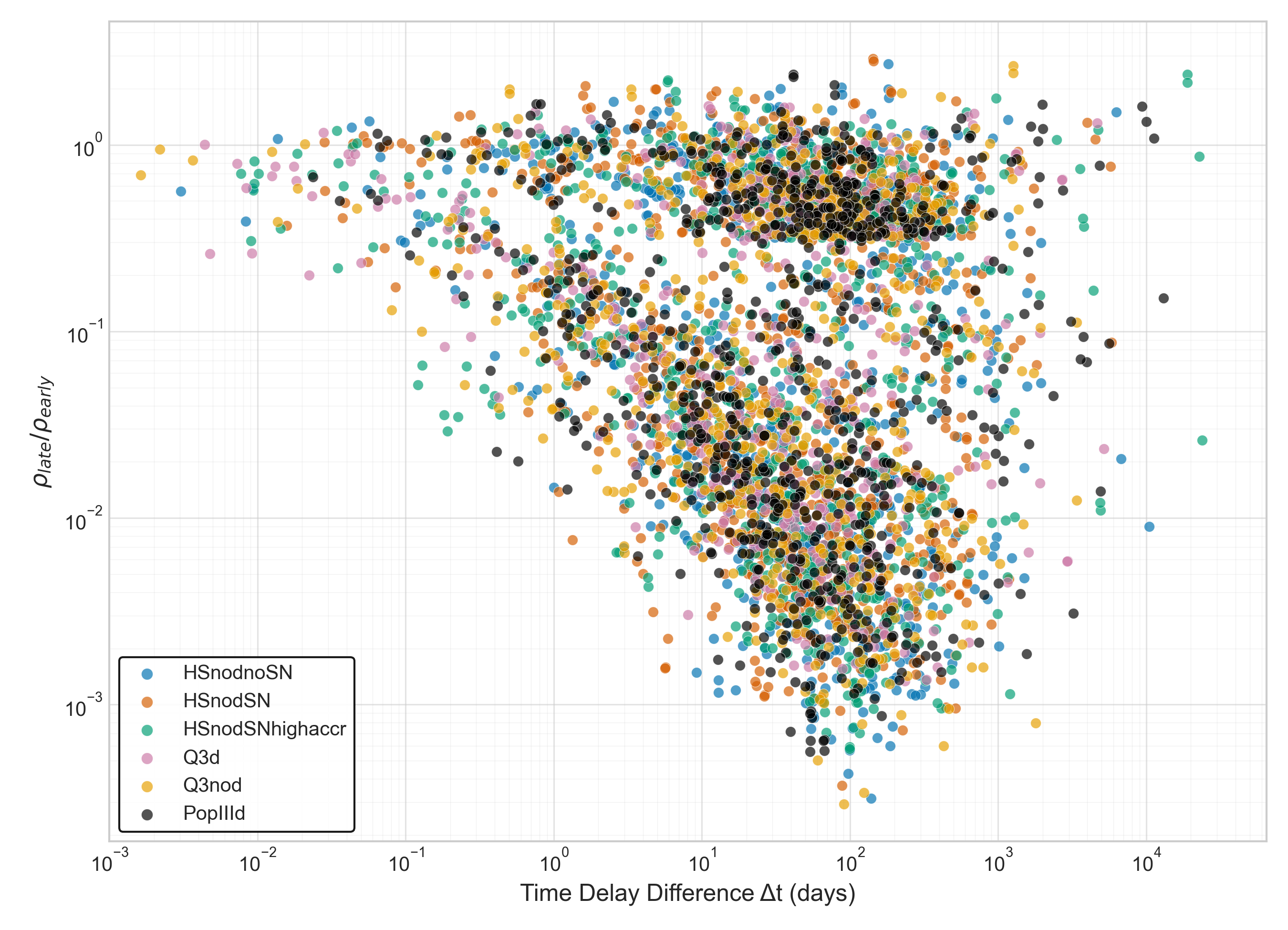}
    \caption{Time delay difference distribution combined with SNR ratio distribution in the 100,000 samples in LISA observation. Different colors represent different formation models.}
    \label{Figure:time_ratio_distribution}
\end{figure}

\begin{figure*}
    \centering
    \includegraphics[width=1.0\linewidth]{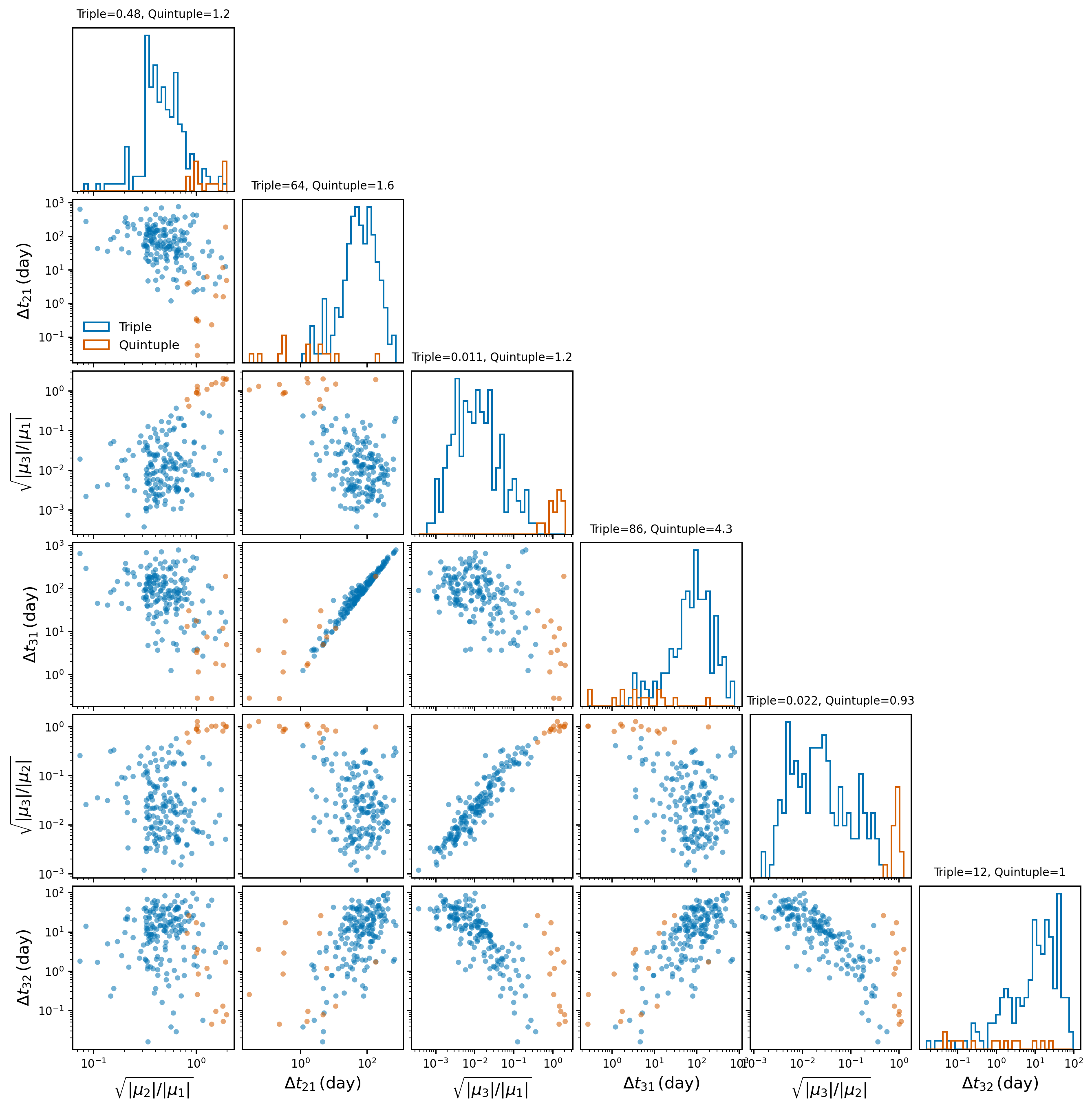}
    \caption{The joint distribution of the time delay difference and the magnification ratio for the model \texttt{HSnodSN} in 100,000 samples of LISA observation. The triple and quintuple plots (blue and orange points) represent the distribution of the baseline 3-image and 5-image systems, which are already shown in the Table~\ref{tab:original_compare_overlap}. The value shows in the top of the figure represents the median of each distribution.}
    \label{Figure:time_ratio_distribution_corner}
\end{figure*}

As shown in \citep{li2026mockcatalogsstronglylensed}, for lensed signals without considering signal overlap, we can define $\rho_{\mathrm{late}}$ and $\rho_{\mathrm{early}}$ as the SNR of early and late arriving signals, respectively. Then the ratio $\rho_{\mathrm{late}}/\rho_{\mathrm{early}}$ can be used to construct a Bayes factor for identifying lensed events. In this case, we present the distribution of $\rho_{\mathrm{late}}/\rho_{\mathrm{early}}$ together with the time-delay differences between lensed images, $\Delta t_{ij}$, in Figure~\ref{Figure:time_ratio_distribution}. For the LISA case, however, the overlap of lensed signals must be taken into account. For each overlapped signal, as shown in Eq.~\ref{lensed_waveform}, the quantities $\sqrt{|\mu_i|/|\mu_j|}$ and $\Delta t_{ij}$ become the key parameters governing the lensed waveform. Consequently, the ratio $\rho_{\mathrm{late}}/\rho_{\mathrm{early}}$ is no longer an appropriate quantity for identifying lensed signals. Instead, we present the joint distribution of $\sqrt{|\mu_i|/|\mu_j|}$ and $\Delta t_{ij}$ in Figure~\ref{Figure:time_ratio_distribution_corner}. In addition, the baseline 3-image and 5-image systems play different roles in the parameter-estimation process. We therefore distinguish them using different colors in the corresponding figure. Here we show the results for the \texttt{HSnodSN} model as a representative example, analogous plots for the other models are available in the GitHub repository mentioned above.

In the end of this section, we summarize the results comparing baseline scenario to the scenario when we consider the overlapping of the lensed signals in Table~\ref{tab:original_compare_overlap}.

\begin{deluxetable*}{lccccccccccccc}
\tablecaption{Comparison of the number of lensed events with different numbers of detectable signals between baseline scenario and the scenario when the overlapping lensed signals are considered for different formation models in 100,000 samples of LISA detection.}
\tabletypesize{\scriptsize}
\label{tab:original_compare_overlap}

\tablehead{
\colhead{} & \colhead{} &
\multicolumn{4}{c}{Baseline 3-image system} & 
\multicolumn{6}{c}{Baseline 5-image system} &
\multicolumn{2}{c}{Baseline single-image system} \\
\cmidrule(lr){3-6} \cmidrule(lr){7-12} \cmidrule(lr){13-14}
\colhead{Source Model} & \colhead{Scenario} & 
\colhead{$N_{0}^{(3)}$} & \colhead{$N_{1}^{(3)}$} & \colhead{$N_{2}^{(3)}$} & \colhead{$N_{3}^{(3)}$} & 
\colhead{$N_{0}^{(5)}$} & \colhead{$N_{1}^{(5)}$} & \colhead{$N_{2}^{(5)}$} & \colhead{$N_{3}^{(5)}$} & \colhead{$N_{4}^{(5)}$} & \colhead{$N_{5}^{(5)}$} & 
\colhead{$N_{r_{\mathrm{max}}>8\sqrt{3}/\rho_0}$} & \colhead{$N_{r_{\mathrm{max}}>8\sqrt{10}/\rho_0}$}
}

\startdata
\multirow{2}{*}{HSnodnoSN} & Baseline & 3 & 30 & 181 & 24 & 0 & 1 & 0 & 2 & 12 & 2 & 66 & 58 \\
          & Overlap  & 3 & 183 & 50 & 2 & 0 & 16 & 1 & 0 & 0 & 0 & 66 & 58\\
\multirow{2}{*}{HSnodSN} & Baseline & 1 & 24 & 205 & 16 & 0 & 0 & 0 & 2 & 14 & 6 & 62 & 53 \\
        & Overlap  & 1 & 201 & 43 & 1 & 0 & 14 & 2 & 4 & 1 & 1 & 62 & 53 \\
\multirow{2}{*}{HSnodSNhighaccr} & Baseline & 1 & 12 & 183 & 32 & 0 & 0 & 0 & 1 & 20 & 0 & 74 & 68 \\
                & Overlap  & 1 & 173 & 54 & 0 & 0 & 18 & 2 & 1 & 0 & 0 & 74 & 68 \\
\multirow{2}{*}{PopIIId} & Baseline & 122 & 63 & 31 & 2 & 5 & 4 & 2 & 5 & 5 & 0 & 11 & 4 \\
        & Overlap  & 99 & 119 & 0 & 0 & 2 & 19 & 0 & 0 & 0 & 0 & 11 & 4 \\
\multirow{2}{*}{Q3d} & Baseline & 0 & 0 & 56 & 113 & 0 & 0 & 0 & 0 & 2 & 6 & 37 & 37 \\
    & Overlap  & 0 & 72 & 64 & 33 & 0 & 2 & 1 & 0 & 4 & 1 & 37 & 37 \\
\multirow{2}{*}{Q3nod} & Baseline & 0 & 7 & 136 & 51 & 0 & 0 & 0 & 0 & 11 & 3 & 45 & 45 \\
      & Overlap  & 0 & 124 & 68 & 2 & 0 & 9 & 1 & 2 & 2 & 0 & 45 & 45 \\
\enddata
\tablecomments{Baseline 3/5/single-image system are dirctly calculated from out lensing model without considering the SNR threshold.Symbol $N_{i}^{(j)}$ denote the number of events with $i$ detectable signals among the baseline $j$-image system in two scenarios, while $N_{r_{\mathrm{max}}>8\sqrt{3}/\rho_0}$ and $N_{r_{\mathrm{max}}>8\sqrt{10}/\rho_0}$ represent the number of events with max magnification factor greater than $8\sqrt{3}/\rho_0$ and $8\sqrt{10}/\rho_0$, with $\rho_0$ represents the unlensed SNR, respectively.}
\end{deluxetable*}

\begin{deluxetable*}{lccc}
\tablecaption{Intrinsic parameters of the BBH, BNS, and NSBH populations}
\tabletypesize{\scriptsize}
\label{tab:bbh_intrinsic_parameters}
\tablehead{\colhead{Parameter} & \colhead{BBH} & \colhead{BNS} & \colhead{NSBH}}
\startdata
$m_1$ & Broken power-law with two peaks; $m_1 \in [5,100]\,M_\odot$ & Peak model; $m_1 \in [1.0,2.4]\,M_\odot$ & Truncated power law; $m_1 \in [2,20]\,M_\odot$ \\
$q$ & Power-law with smoothing; $q \in [5.0/m_1,1]$ & Power law; $q \in [1.0/m_1,1]$ & Truncated Gaussian; $q \in [1.0/m_{\mathrm{BH}},q_{\mathrm{max}}]$ \\
$z$ & \multicolumn{3}{c}{Following \citep{2013ApJ...779...72D}; $z \in [0,21.5]$} \\
$\chi$ & Truncated normal; $\chi \in (0,1)$ & Uniform; $\chi \in (0,0.05)$ & $\chi_{\mathrm{BH}} \sim \mathrm{Beta}(0,1),\ \chi_{\mathrm{NS}} \sim \mathcal{U}(0,0.05)$ \\
\enddata
\tablecomments{The BBH population adopts the fiducial \textit{Broken Power Law + 2 Peaks} mass model (Appendix B.5.1) and the \textit{Gaussian Component Spins} model (Appendix B.3) from \citep{Abac_2025_GWTC4}. For BNS events, the primary mass follows the \textit{Peak model} (Appendix B.2), with a low-spin assumption ($\chi \leq 0.05$) \citep{Abac_2025_GWTC4}. The NSBH population is modeled using the hierarchical framework of \citep{2023MNRAS.518.5298B}, in which black hole spins follow a Beta distribution and mass ratios follow a truncated Gaussian.}
\end{deluxetable*}

\section{Detector: DECIGO}\label{sec:decigo}
In this section, we will consider the case of DECIGO, which is a proposed space-borne gravitational wave detector with sensitivity in the decihertz frequency band (0.1-100 Hz). Same as the LISA case in Section ~3, we will first describe the source model and then present the resulting mock catalog of strongly lensed GWs detectable by DECIGO. 

\subsection{Source Model}
Several studies have investigated different science targets for the DECIGO detector. For example, \cite{2025arXiv251109107S} combines DECIGO with the Einstein Telescope (ET) to trace the evolution of BNS systems from the early inspiral to the pre-merger phase. In this section, instead, we focus on stellar-mass BBHs, BNSs, and NSBHs, aiming to estimate how many lensed events can be detected by DECIGO alone, without the support of ground-based detectors.

The intrinsic parameters of the BBH, BNS, and NSBH populations are adopted from our previous work \citep{li2026mockcatalogsstronglylensed}, as summarized in Table~\ref{tab:bbh_intrinsic_parameters}. Further details can be found in the above mentioned paper.

\begin{figure}
    \centering
    \includegraphics[width=1.0\linewidth]{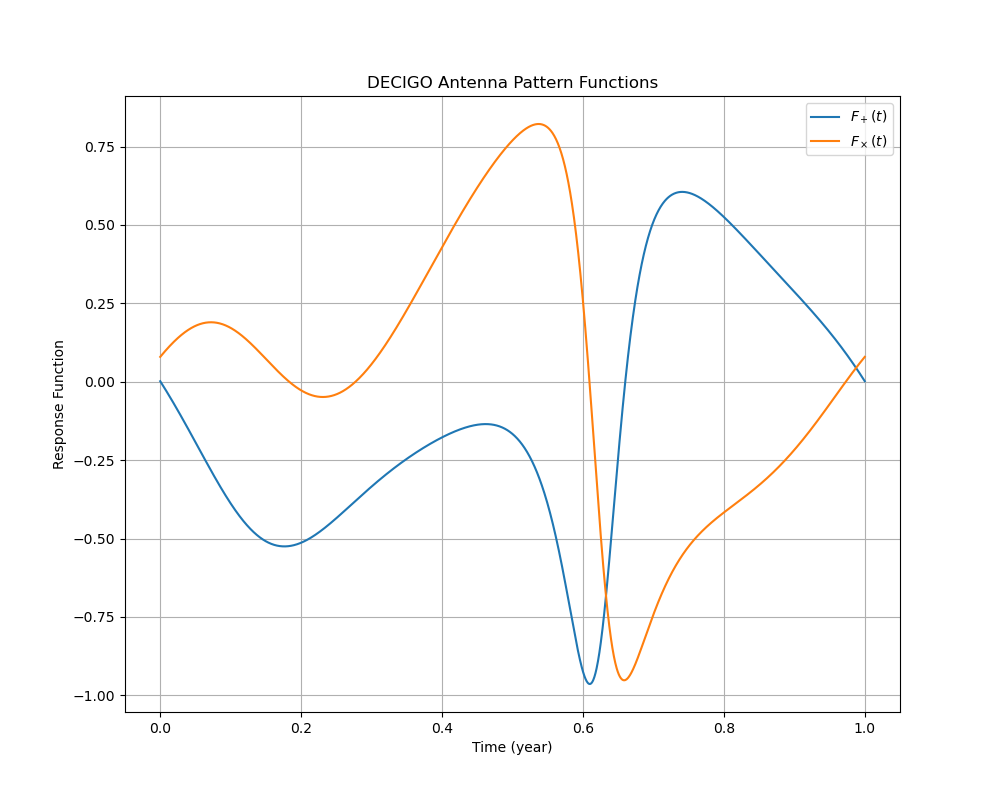}
    \caption{The antenna pattern function of DECIGO in 1yr observation, with the source position fixed at $\bar{\theta}_S = \bar{\phi}_S =\pi/4$ and the orientation of the orbital angular momentum fixed at $\hat{\theta}_L = \hat{\phi}_L = 0$.}
    \label{response}
\end{figure}

\begin{deluxetable*}{lcc}
\tablecaption{Extrinsic parameters of the BBH population for DECIGO case}
\label{tab:decigo_extrinsic_parameters}
\tablehead{
\colhead{Parameter} & \colhead{Physical Meaning} & \colhead{Distribution}
}
\startdata
$(\bar{\theta}_S, \bar{\phi}_S)$ & Source sky position (heliocentric frame) 
& \textbf{Isotropic}, $\cos\bar{\theta}_S \sim U(-1,1)$, $\bar{\phi}_S \sim U(0,2\pi)$ \\
$(\bar{\theta}_L, \bar{\phi}_L)$ & Orientation of angular momentum 
& \textbf{Isotropic}, $\cos\bar{\theta}_L \sim U(-1,1)$, $\bar{\phi}_L \sim U(0,2\pi)$ \\
$\phi_c$ & Coalescence phase 
& \textbf{Uniform}, $\phi_c \sim U(0,2\pi)$ \\
$t_c$ & Coalescence time 
& \textbf{Uniform}, $t_c \sim U(0, 1yr)$ \\ 
\enddata
\end{deluxetable*}
Before we start discussing the extrinsic parameters we need to clarify the response function and the power spectral density (PSD) of DECIGO. Since there is no dedicated software available which can be used to calculate the SNR for DECIGO, we will use the response function given in \cite{2011PhRvD..83d4011Y}. Its analytical form is:
\begin{equation}
    h(t) = F_+(\bar{\theta}_S, \bar{\phi}_S, \hat{\theta}_L, \hat{\phi}_L) h_+(t) + F_{\times}(\bar{\theta}_S, \bar{\phi}_S, \hat{\theta}_L, \hat{\phi}_L) h_{\times}(t)\,,
\end{equation}
where
\begin{equation}
\begin{aligned}
F_+(\theta_S, \phi_S, \psi_S) 
&= \frac{1}{2}(1+\cos^2\theta_S)\cos 2\phi_S \cos 2\psi_S \\
&\quad - \cos\theta_S \sin 2\phi_S \sin 2\psi_S \,,\\
F_{\times}(\theta_S, \phi_S, \psi_S) 
&= \frac{1}{2}(1+\cos^2\theta_S)\cos 2\phi_S \sin 2\psi_S \\
&\quad + \cos\theta_S \sin 2\phi_S \cos 2\psi_S \,.
\end{aligned}
\end{equation}
are antenna patterns. The $(\bar{\theta}_S, \bar{\phi}_S)$ denote the source position in the heliocentric coordinate system and $(\hat{\theta}_L, \hat{\phi}_L)$ denote the orientation of the orbital angular momentum while the $(\theta_S, \phi_S, \psi_S)$ represent the source position in the detector frame and the polarization angle. The transformation between the two coordinate systems is given by:
\begin{equation}
    \label{coordinate_trans}
    \begin{aligned}
        \theta_S(t) &= \arccos\left[\frac{1}{2}\cos\bar{\theta}_S - \frac{\sqrt{3}}{2}\sin\bar{\theta}_S \cos(\hat{\phi}(t) - \bar{\phi}_S)\right]\,,\\
        \phi_S(t) &= \frac{\pi}{12}+\arctan\left[\frac{\sqrt{3}\cos\bar{\theta}_S +\sin\bar{\theta}_S\cos(\hat{\phi}(t) - \bar{\phi}_S)}{2\sin\bar{\theta}_S\sin(\hat{\phi}(t)-\bar{\phi}_S)}\right]\,,\\
        \psi_S &= \arctan\left(\frac{a}{b}\right)\,,
    \end{aligned}
\end{equation}
where
\begin{equation}
    \label{ab}
\begin{aligned}
a&=  \frac{1}{2} \cos \hat{\theta}_L-\frac{\sqrt{3}}{2} \sin \hat{\theta}_L \cos \left[\hat{\phi}(t)-\hat{\phi}_L\right] \\
& -\cos \hat{\theta}_L \cos ^2 \bar{\theta}_S-\sin \hat{\theta}_L \sin \bar{\theta}_S \cos \left(\hat{\phi}_L-\bar{\phi}_S\right) \,,\\
b&= \frac{1}{2} \sin \hat{\theta}_L \sin \bar{\theta}_S \sin \left(\hat{\phi}_L-\bar{\phi}_S\right) \\
& -\frac{\sqrt{3}}{2} \cos \hat{\phi}(t)\left[\cos \hat{\theta}_L \sin \bar{\theta}_S \sin \bar{\phi}_S-\cos \bar{\theta}_S \sin \hat{\theta}_L \sin \hat{\phi}_L\right] \\
& -\frac{\sqrt{3}}{2} \sin \hat{\phi}(t)\left[\cos \bar{\theta}_S \sin \hat{\theta}_L \cos \hat{\phi}_L-\cos \hat{\theta}_L \sin \bar{\theta}_S \cos \bar{\phi}_S\right]\,,
\end{aligned}
\end{equation}
and $\hat{\phi}(t)=2\pi t/T$ with $T=1\,\mathrm{yr}$ being the orbital period of DECIGO around the Sun while $\bar{\theta}(t) = \pi/2$. The value of the response function calculated by the above formula is shown in Figure~\ref{response}. And according to \cite{2021Galax...9...14I}, the DECIGO mission is designed with spacecraft clusters in heliocentric, Earth-like orbits around the Sun. As for the noise power spectral density (PSD), we will use the one given in \cite{2011PhRvD..83d4011Y}, the analytical form is:
\begin{equation}
    \label{DECIGO_psd}
    \begin{aligned}
S_n(f)= & 10^{-48} \times\left[7.05\left(1+\frac{f^2}{f_p^2}\right)+4.80 \times 10^{-3} \times\right. \\
& \left.\frac{f^{-4}}{1+\left(f / f_p\right)^2}+5.33 \times 10^{-4} f^{-4}\right] \mathrm{Hz}^{-1}\,,
\end{aligned}
\end{equation}
where $f_p = 7.36\,\mathrm{Hz}$. 
As mentioned in the previous section, in the case of LISA we considered the confusion noise from unresolved binary systems in our Galaxy. Similar effect should be considered for DECIGO, too as shown in \cite{Piorkowska2021}. Now, however, the unresolved sources are BBH, BNS and BHNS systems (collectively called double compact objects DCOs) at cosmological distances. It is common to characterize the magnitude of a stochastic GW background by its energy density per logarithmic frequency interval $\Omega^{DCO}_{GW} = \frac{1}{\rho_{cr}} \frac{d \rho^{DCO}_{GW}}{d \ln{f}}$ normalized by the critical density of the Universe $\rho_{cr} = \frac{3 H_0^2}{8 \pi G}$. Energy density parameter $\Omega^{DCO}_{GW}$ corresponding to the unresolved signals from the DCO systems \citep{Piorkowska2021} can be expressed as 
\begin{align}
    \Omega^{\mathrm{DCO}}_{\mathrm{GW}} = &\Omega^{\mathrm{DCO}}_{0}\left(\frac{H_0}{70\,\mathrm{km\,s^{-1}\,Mpc^{-1}}}\right)^{-3}\\&\times\left(\frac{\mathcal{M}}{M_{\mathrm{DCO}}}\right)^{5/3}\left(\frac{f}{1\,\,\mathrm{Hz}}\right)^{2/3},
\end{align}
where $\Omega^{\mathrm{DCO}}_{0} = \frac{8 (\pi G \mathcal{M_{\mathrm{DCO}}})^{5/3}}{9 c^2 H_0^2} \int_0^{\infty} \frac{{\dot n}(z)}{(1+z)^{4/3} H(z)} dz, $ ${\dot n}(z)$ is the DCO merger rate per proper time per comoving volume at redshift $z$ and $\mathcal{M_{\mathrm{DCO}}}$ is the median value of the DCO considered. $H(z)$ is the expansion rate in $\Lambda$CDM model considered in this paper. The source population of DCOs studied in this paper, following our previous work \citep{li2026mockcatalogsstronglylensed} is the same as considered in \citep{Piorkowska2021} and derived from the same population-synthesis model. Therefore, we may use the values ($\mathcal{M_{\mathrm{NSNS}}};\,\mathcal{M_{\mathrm{BHNS}}};\,\mathcal{M_{\mathrm{BHBH}}}$) = $(1.22;\,6.09;\,24.5)M_{\odot}$, and the characteristic quantity, $\Omega_0^{\mathrm{DCO}}$, as reported in Table~1 in \cite{Piorkowska2021}. 
Consequently, one can express the confusion noise power spectrum as :
\begin{equation}
    \label{dcosp}
    S_{n}^{\mathrm{DCO}}(f) = \frac{4}{\pi}f^{-3}\rho_{\mathrm{cr}}\Omega_{\mathrm{GW}}^{\mathrm{DCO}}.
\end{equation}
Hence, accounting for the confusion noise we use the modified noise power spectrum $S_{n}^{eff} (f)= S_{n}(f) + S_{n}^{\mathrm{DCO}}(f)$.

Using the detector response function and the efficient noise power spectral density (PSD), we compute the SNR for each event in the same manner as in the LISA case described in Section~3. A summary of all extrinsic parameters and their distributions is provided in Table~\ref{tab:decigo_extrinsic_parameters}. Here we adopt the \texttt{IMRPhenomXPHM} waveform model, which introduces four additional parameters related to the spin orientations, $(\theta_1, \theta_2, \phi_{12}, \phi_{JL})$. In total, 15 parameters are used to compute the SNR in the DECIGO case.

\subsection{Lensed GW catalog: DECIGO Case}
\begin{figure*}
    \centering
    \includegraphics[width=1.0\linewidth]{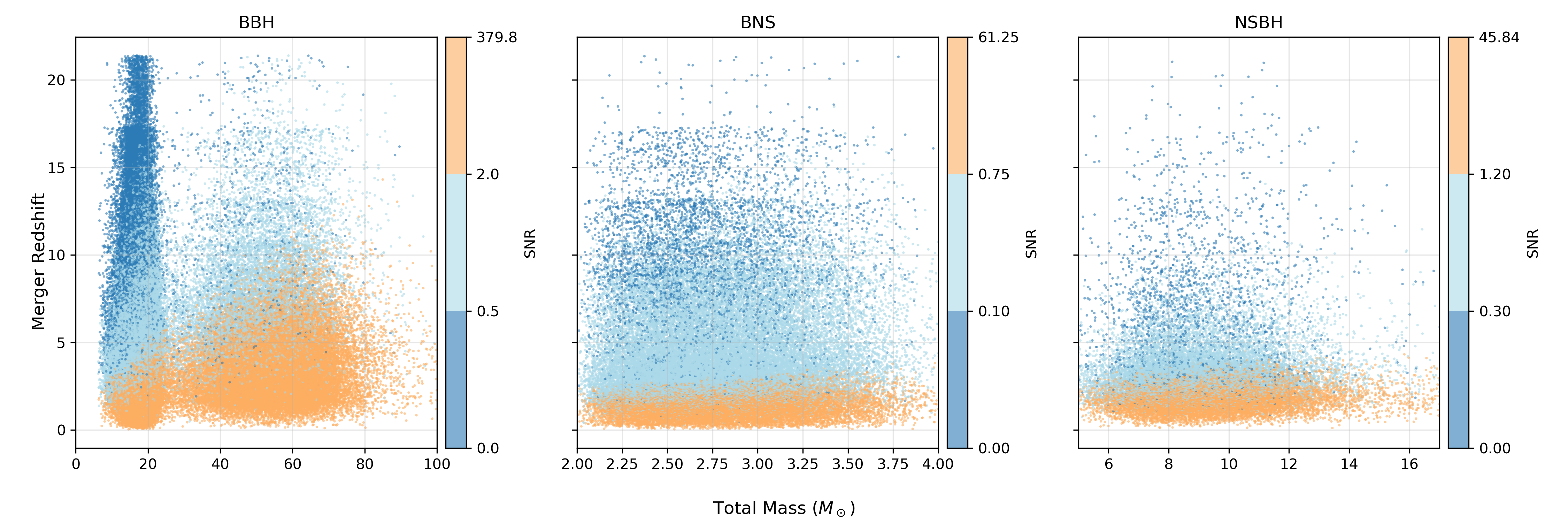}
    \caption{SNR distributions of the intrinsic populations of BBHs, BNSs, and NSBHs for the DECIGO case over a one-year observation period. In each subplot, the SNR distribution is divided into three regimes to distinguish different SNR ranges.}
    \label{Figure: decigo_snr_distribution}
\end{figure*}

Using the distributions of intrinsic and extrinsic parameters, we compute the SNR for each event. We employ \texttt{bilby}\footnote{\url{https://bilby-dev.github.io/bilby/}} {\citep{colm_talbot_2025_17533961,bilby_paper}} to generate the waveforms and evaluate the SNR. The resulting SNR distribution is shown in Figure~\ref{Figure: decigo_snr_distribution}. 

We adopt the CBC rates from our previous work \citep{li2026mockcatalogsstronglylensed} to estimate the number of events for each source type in a one-year DECIGO observation; the corresponding values are listed in Table~\ref{tab:decigo_event_number}. We find that the number of detectable events with $\rho_0 > 8$ is substantial for all three source types, with BBHs yielding the largest number of detections, followed by NSBHs and BNSs. The SNR distributions indicate that the confusion noise significantly declines the number of detectable lensed signals.

\begin{deluxetable}{lcc}
\tablecaption{Intrinsic and detectable event numbers of BBHs, BNSs, and NSBHs for DECIGO case in 1-year observation}
\label{tab:decigo_event_number}
\tablehead{
\colhead{Source Type} &
\colhead{Intrinsic number} &
\colhead{Detectable number ($\rho_0 > 8$)} 
}
\startdata
BBH & 404698 & 15953  \\
BNS & 50661 & 367 \\
NSBH & 30260 & 251 \\
\enddata
\end{deluxetable}

\begin{figure}
    \centering
    \includegraphics[width=1.0\linewidth]{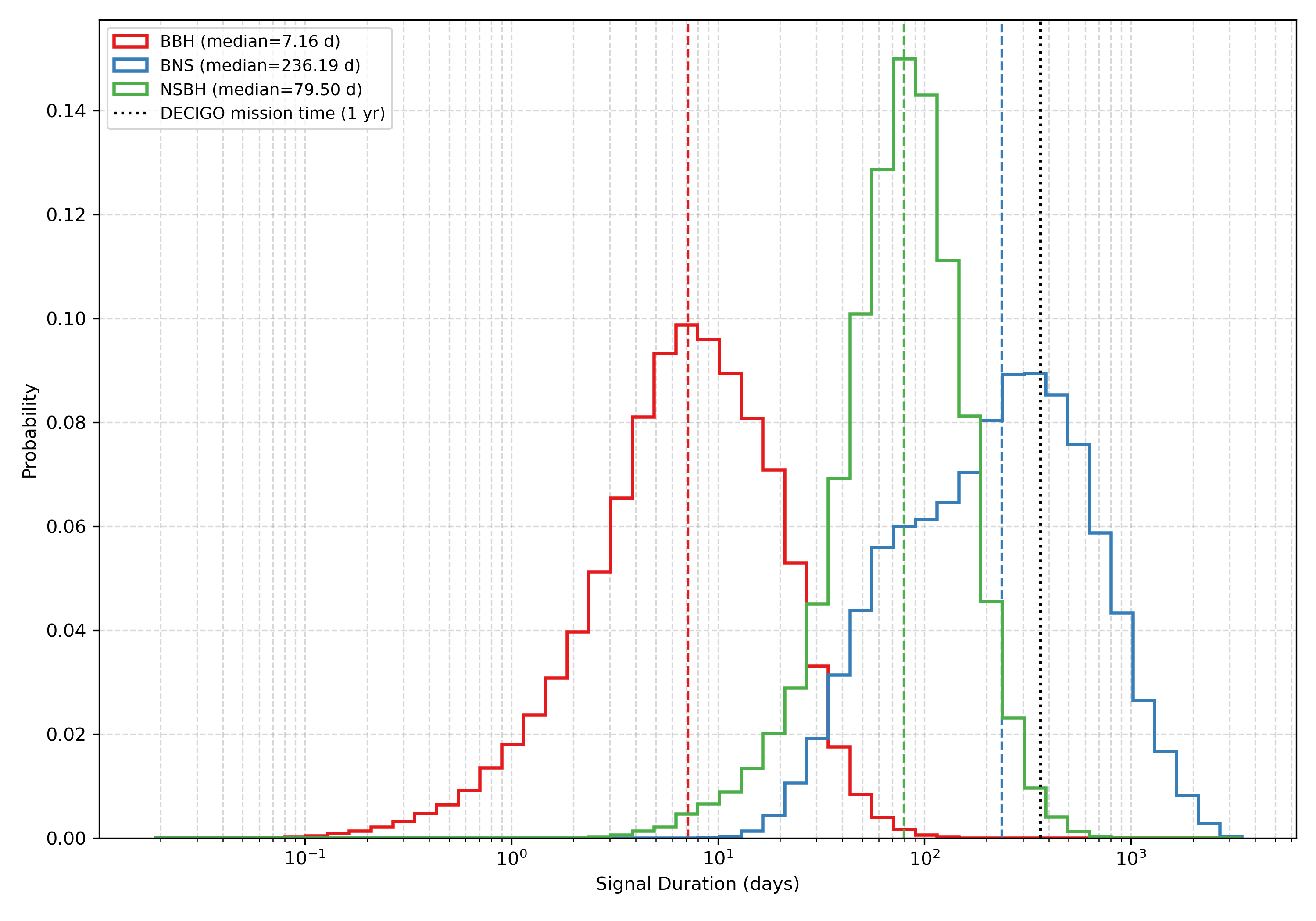}
    \caption{Distribution of the gravitational-wave signal duration for BBHs, BNSs, and NSBHs during a one-year DECIGO mission. The durations are calculated for all events in each source type.}
    \label{Figure:decigo_duration_time_distribution}
\end{figure}

Before applying the lensing model to the intrinsic population, we must first estimate the signal duration for each event. Unlike the LISA case, signal in DECIGO band will be always in the inspiral phase, so the signal duration can be calculated using the following formula:
\begin{equation}
    \label{delta_t_obs_modify}
\Delta t_{\mathrm{obs}}=
\frac{5}{256}
\left(\frac{G\mathcal{M}_z}{c^3}\right)^{-5/3}
\pi^{-8/3}
\left(
f_{\mathrm{low}}^{-8/3}
-
f_{\mathrm{high}}^{-8/3}
\right)\,,
\end{equation}
where $f_{\mathrm{low}} = 0.1\,\mathrm{Hz}$ and $f_{\mathrm{high}} = 100\,\mathrm{Hz}$. The resulting distribution of signal durations is shown in Figure~\ref{Figure:decigo_duration_time_distribution}.

We find that the signal durations for stellar-mass BBHs, BNSs, and NSBHs in the DECIGO band are generally much shorter than those of MBHBs in the LISA band. This implies that, for DECIGO, signal overlap induced by lensing time delays is less likely to occur compared to LISA. Nevertheless, it remains important to account for this effect in cases involving long-duration signals and short lensing time delays, particularly for BNSs, which can exhibit longer durations than BBHs and NSBHs. 

\begin{figure}
    \centering
    \includegraphics[width=1.0\linewidth]{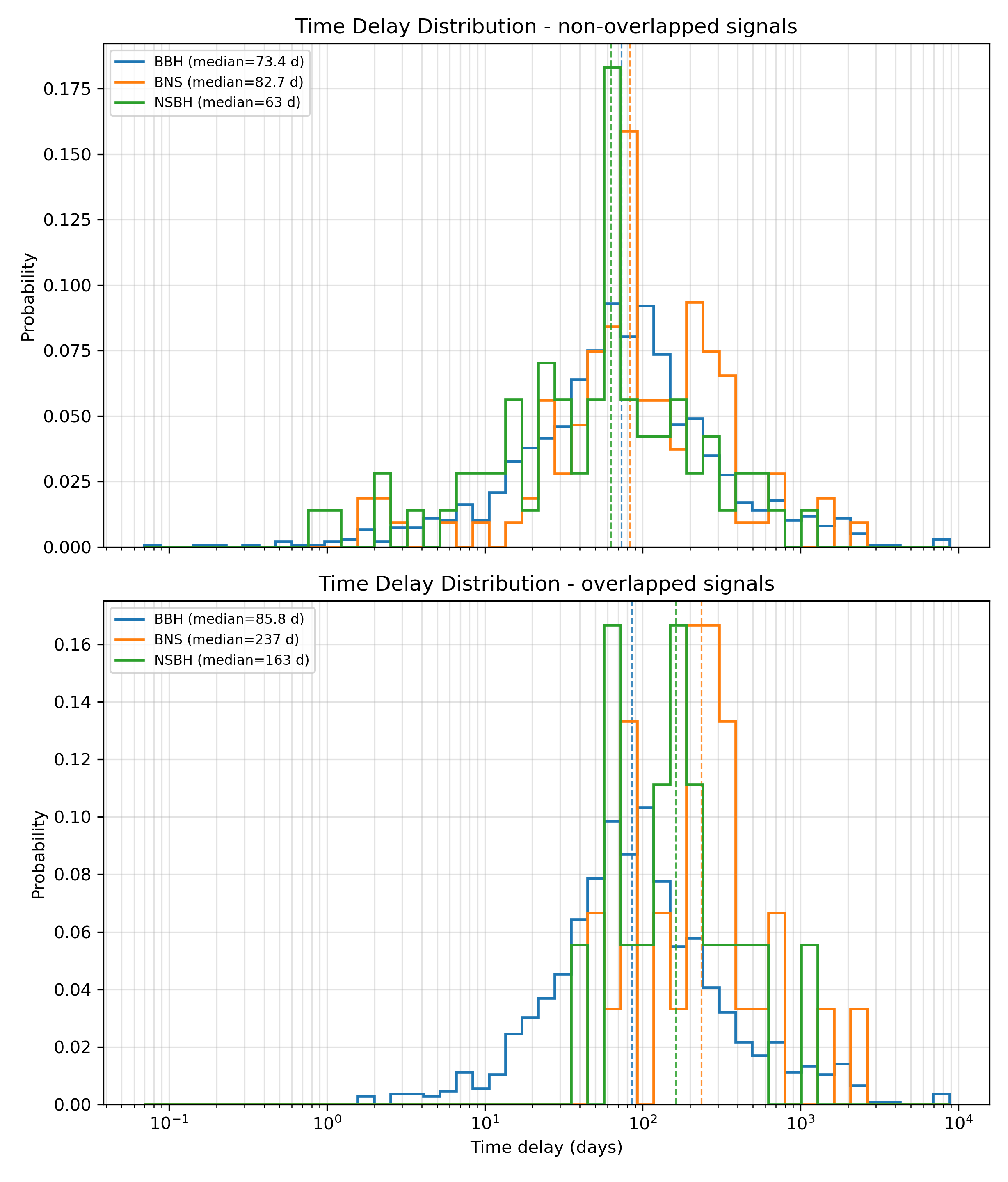}
    \caption{Time delay distribution of the lensed events for BBHs, BNSs, and NSBHs in the DECIGO band. The upper panel shows the time delay distribution without considering the overlapping of lensed signals, while the lower panel shows the distribution when we consider the overlapping.}
    \label{Figure:decigo_time_delay_distribution}
\end{figure}

The relationship between signal duration and time delay can be inferred by comparing Figure~\ref{Figure:decigo_duration_time_distribution} with Figure~\ref{Figure:decigo_time_delay_distribution}. Accordingly, we adopt the same strategy as in the LISA case to compute the SNR of lensed events, incorporating the potential overlap of lensed signals based on their durations and lensing time delays.

The number of lensed events without imposing an SNR threshold for each source type is presented in Table~\ref{tab:decigo_lensed_event_number_original}. The lensed SNR is computed using Eqs.~(\ref{snr_fullwaveform}--\ref{lensed_waveform}). We find a lensing probability of approximately $0.15\%$, along with the number of lensed events corresponding to different numbers of observable signals, both with and without accounting for signal overlap. 

Similar to the LISA case, although many lensing systems formally produce multiple images, the number of independent signals is significantly reduced when overlap between lensed signals is taken into account. This again demonstrates that signal overlap is an important factor that must be considered when analyzing strongly lensed GW events in the DECIGO band, particularly for BNSs, which can have relatively long signal durations. 

\begin{deluxetable*}{lcccccc} 
    \tablecaption{Baseline lensed event number of BBHs, BNSs, and NSBHs without considering the threshold for DECIGO case in 1-year observation} 
    \label{tab:decigo_lensed_event_number_original} 
    \tablehead{ \colhead{Source type} & \colhead{Total number} & \colhead{1-signal} & \colhead{2-signal} & \colhead{3-signal} & \colhead{4-signal} & \colhead{5-signal} } 
    \startdata 
    BBH & 784 & 127(161) & 38(233) & 582(363) &4(10) & 33(17) \\ 
    BNS & 68 & 13(42) & 6(22) & 47(4) &1(0) & 1(0) \\ 
    NSBH & 41 & 10 (26)& 1(12) & 25(3) &0(0) & 5(0) \\
    \enddata 
    \tablecomments{The numbers in the parentheses represents the number of lensed events when the
signal overlap is considered.}
\end{deluxetable*}

\begin{deluxetable*}{lcccccc} 
    \tablecaption{Statistical results of detected lensed events for BBHs, BNSs, and NSBHs for DECIGO case in 1-year observation} 
    \tablehead{ \colhead{Source type} & \colhead{Detected number} & \colhead{$N_{1}$} & \colhead{$N_{2}$} & \colhead{$N_{sub}$} & \colhead{$N_{r_{\mathrm{max}}>3}$} & \colhead{$N_{r_{\mathrm{max}}>10}$} } 
    \startdata 
    BBH           & 44 & 37 & 7 & 3 & 21 & 0 \\
    BNS           & 0 & 0 & 0 & 0 & 1 & 0\\
    NSBH         & 1 & 1 & 0 & 0 & 5 & 0 \\
    \enddata 
    \tablecomments{All the symbols are considered as the same as the defination in Table~\ref{tab:catalog_lensed_event_number_final}.}
    \label{tab:catalog_DECIGO_lensed_event_number_final} 
\end{deluxetable*}

For the detected lensed events, we summarize the statistical results for each source type when signal overlap is taken into account in Table~\ref{tab:catalog_DECIGO_lensed_event_number_final}. The total number of detected lensed events ranges from 0 for BNSs to 44 for BBHs. Similar to the LISA case, the majority of events produce a single detectable signal, while a non-negligible fraction yield multiple detectable signals due to lensing. The number of events lensed by subhalos is also significant, indicating that substructure lensing can play an important role in shaping the observed properties of lensed GWs in the DECIGO band.

\begin{figure*}
    \centering
    \includegraphics[width=1.0\linewidth]{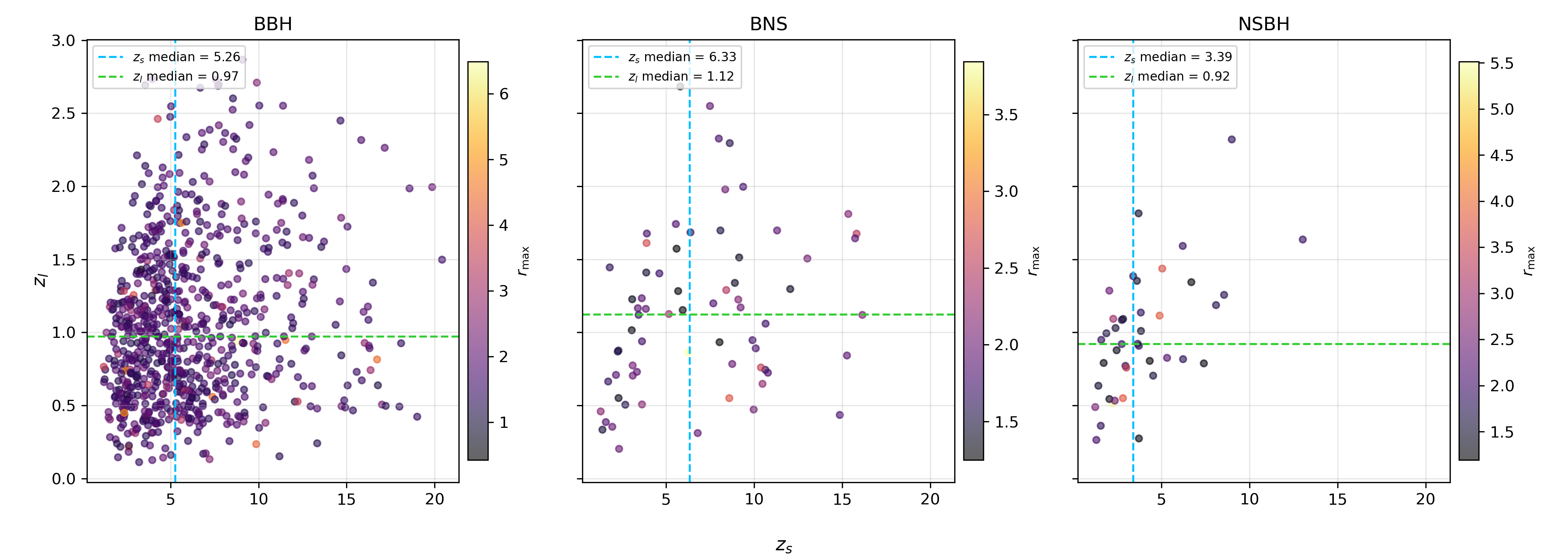}
    \caption{The distribution of lens redshift $z_l$ and the source redshift $z_s$ for all samples in 1-yr observation with the scatter plot of the max SNR ratio $r_{\text{max}}$ for BBHs, BNSs, and NSBHs in the DECIGO band. The color bar represents the value of $r_{\mathrm{max}}$ while the blue and green dashed lines represent the median of the source redshift and lens redshift, respectively.}
    \label{Figure:decigo_snr_distribution_lensed}
\end{figure*}

Furthermore, a significant fraction of lensed events have maximum magnification factors greater than 3 ($r_{\mathrm{max}} > 3$), while no events exhibit extreme magnifications ($r_{\mathrm{max}} > 10$) in the DECIGO case, in contrast to the LISA results. This difference may be due to the different source populations and lensing configurations probed by DECIGO compared to LISA. The distribution of the maximum magnification $r_{\mathrm{max}}$ for each source type is shown in Figure~\ref{Figure:decigo_snr_distribution_lensed}.

\begin{figure}
    \includegraphics[width=1.0\linewidth]{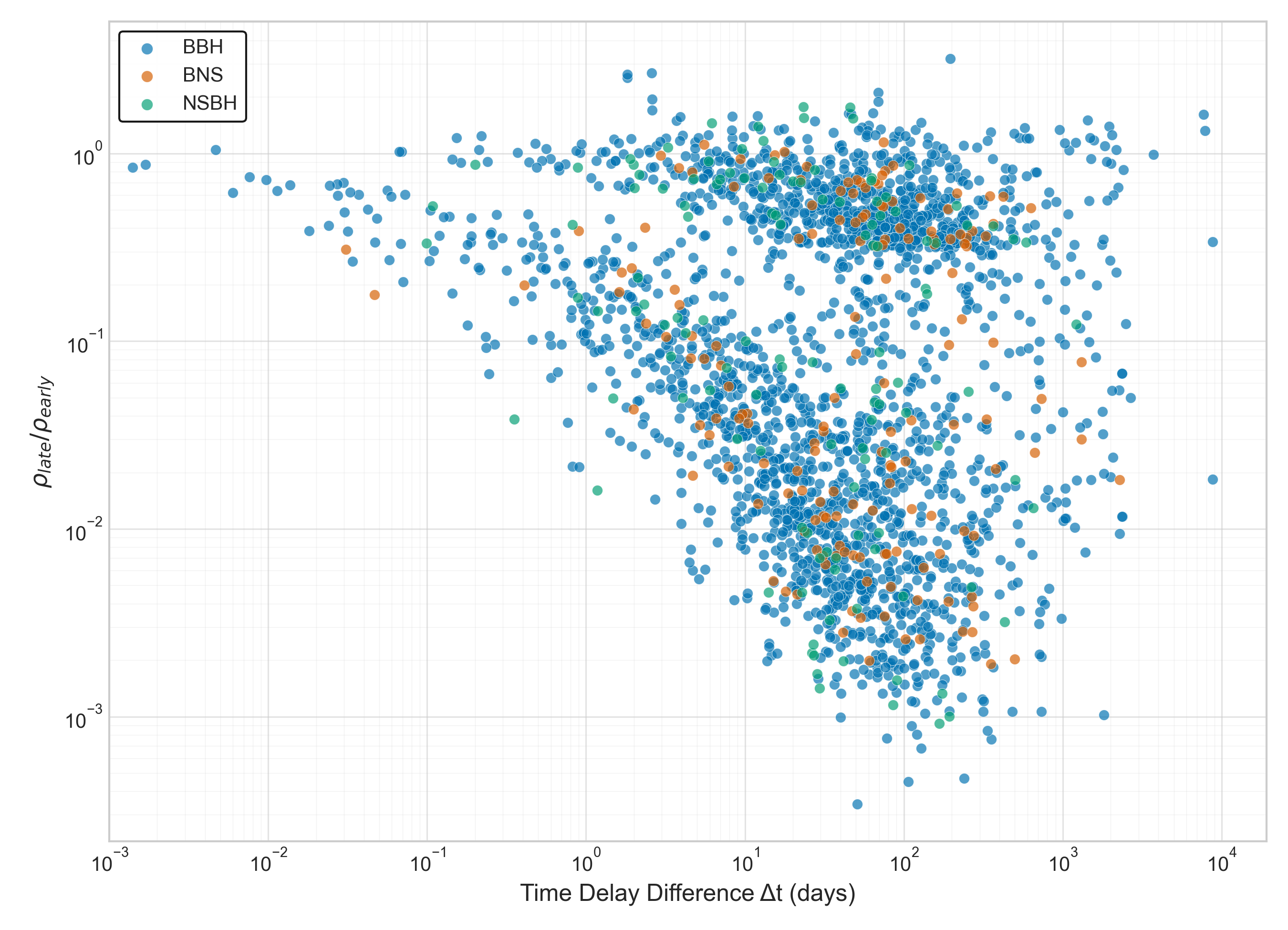}
    \caption{Time delay difference distribution combined with SNR ratio distribution in 1-year observation of DECIGO. Different colors represent different model types.}
    \label{Figure:time_ratio_distribution_DECIGO}
\end{figure}

\begin{figure}
    \includegraphics[width=1.0\linewidth]{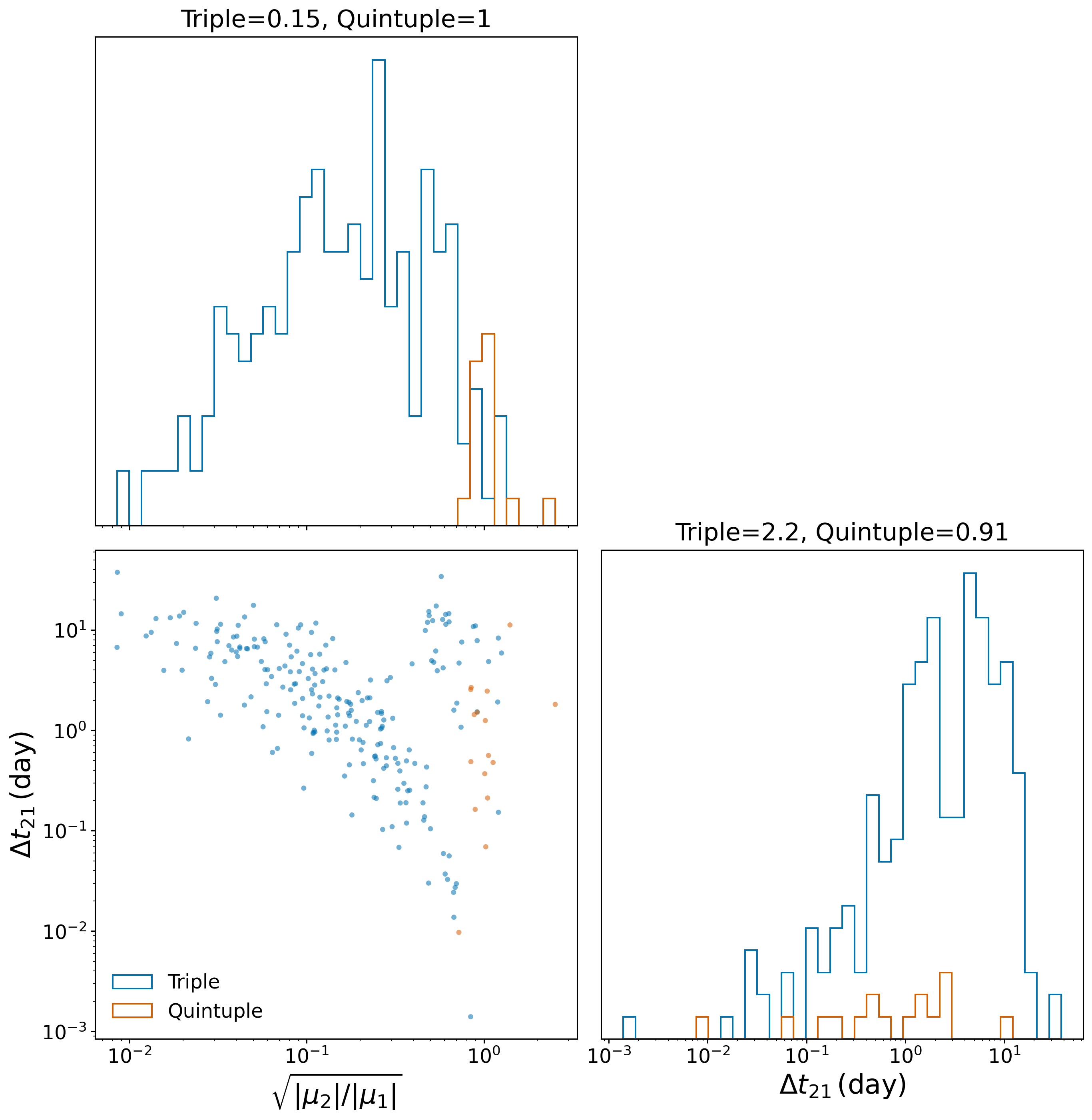}
    \caption{The joint distribution of the time delay difference and the magnification ratio for the source \texttt{BBH} in 1-year DECIGO observation. The triple and quintuple plots (blue and orange points) represent the distribution of the baseline 3-image and 5-image systems, which are already shown in the Table~\ref{tab:original_compare_overlap}. The value shows in the top of the figure represents the median of each distribution.}
    \label{Figure:time_ratio_distribution_corner_DECIGO}
\end{figure}

We also present the joint distributions of the time-delay differences and the SNR ratios for the DECIGO case in Figure~\ref{Figure:time_ratio_distribution_DECIGO}, as well as the joint distributions of the time-delay differences and the magnification ratios in Figure~\ref{Figure:time_ratio_distribution_corner_DECIGO}. Note that, in this case, we only show the relation between $\Delta t_{21}$ and $\sqrt{|\mu_2|/|\mu_1|}$, since configurations involving overlapping signals from three images are much less common for DECIGO. We present only the BBH results here as a representative example; results for the other source classes can also be found in our GitHub repository.

As same as the last section, we also summarize the results comparing baseline scenario to the scenario when we consider the overlapping of the lensed signals in Table~\ref{tab:decigo_original_compare_overlap}.

\section{Conclusions and discussions}\label{sec:conclusion}

\begin{deluxetable*}{lccccccccccccc}
\tablecaption{Comparison of the number of lensed events with different numbers of detectable signals between the Baseline scenario and the scenario when we consider the overlapping of the lensed signals for different source types 1-year observation of DECIGO.}
\tabletypesize{\scriptsize}
\label{tab:decigo_original_compare_overlap}

\tablehead{
\colhead{} & \colhead{} &
\multicolumn{4}{c}{\textbf{Baseline 3-image system}} & 
\multicolumn{6}{c}{\textbf{Baseline 5-image system}} &
\multicolumn{2}{c}{\textbf{Baseline single-image system}}\\
\cmidrule(lr){3-6} \cmidrule(lr){7-12} \cmidrule(lr){13-14}
\colhead{Source Model} & \colhead{Scenario} & \colhead{$N_{0}^{(3)}$} & \colhead{$N_{1}^{(3)}$} & \colhead{$N_{2}^{(3)}$} & \colhead{$N_{3}^{(3)}$} & \colhead{$N_{0}^{(5)}$} & \colhead{$N_{1}^{(5)}$} & \colhead{$N_{2}^{(5)}$} & \colhead{$N_{3}^{(5)}$} & \colhead{$N_{4}^{(5)}$} & \colhead{$N_{5}^{(5)}$} & \colhead{$N_{r_{\mathrm{max}}>8\sqrt{3}/\rho{_0}}$} & \colhead{$N_{r_{\mathrm{max}}>8\sqrt{10}/\rho_0}$}
}
\startdata
\multirow{2}{*}{BBH} & Baseline & 588 & 24 & 8 & 0 & 36 & 0 & 1 & 0 & 0 & 0 & 4 & 1 \\
 & Overlap & 584 & 30 & 6 & 0 & 34 & 2 & 1 & 0 & 0 & 0 & 4 & 1 \\
\multirow{2}{*}{BNS} & Baseline & 53 & 0 & 0 & 0 & 2 & 0 & 0 & 0 & 0 & 0 & 0 & 0 \\
 & Overlap & 53 & 0 & 0 & 0 & 2 & 0 & 0 & 0 & 0 & 0 & 0 & 0 \\
\multirow{2}{*}{NSBH} & Baseline & 25 & 0 & 0 & 0 & 5 & 0 & 0 & 0 & 0 & 0 & 0 & 0 \\
 & Overlap & 24 & 1 & 0 & 0 & 5 & 0 & 0 & 0 & 0 & 0 & 0 & 0 \\
\enddata
\tablecomments{All the symbols are considered as the same as the definition in Table~\ref{tab:original_compare_overlap}.}
\end{deluxetable*}

In the previous sections, we have presented a comprehensive analysis of the expected number of strongly lensed gravitational-wave events from MBHBs in the LISA band, as well as BBHs, BNSs, and NSBHs in the DECIGO band. We have considered various formation models for MBHBs and different source populations for DECIGO, while accounting for the potential overlap of lensed signals induced by lensing time delays. As some of the results may be nontrivial to interpret, we provide in this section a brief summary of the main findings and discuss their implications for future gravitational-wave observations.

\textbf{Detection goals for the LISA and DECIGO cases.} For the LISA case, the primary objective of our simulation is the detection of MBHBs. We adopt six distinct formation models, each characterized by different distributions of mass, redshift, and delay time (see Section~\ref{sec:formation_models}). The largest uncertainties in the predicted lensed event rates arise from differences among these models, with significant variations already evident at the level of the SNR. For example, in terms of the intrinsic SNR, the \texttt{HSnodnoSN} model predicts that the majority of events have $\mathrm{SNR} > 8$, with some reaching values as high as $10^4$. In contrast, the \texttt{PopIIId} model predicts that most events have $\mathrm{SNR} < 8$, indicating a substantial discrepancy between models. A similar level of variation is found in the predicted event rates. The \texttt{HSnodnoSN} model yields 39,364 intrinsic events over a four-year LISA observation, whereas the \texttt{Q3d} model predicts only 74 events. This difference directly propagates to the number of lensed events, even before applying any SNR threshold (see Table~\ref{tab:catalog_lensed_event_number_original}). For the DECIGO case, we focus on the detection of stellar-mass BBHs, BNSs, and NSBHs, with intrinsic parameters adopted from our previous work \citep{li2026mockcatalogsstronglylensed}. Notably, the number of intrinsic BBH events is significantly larger than that of BNSs and NSBHs, leading to a correspondingly higher number of lensed events for BBHs compared to the other two source types (see Table~\ref{tab:decigo_lensed_event_number_original}).

\textbf{Lensing probability and the effect of signal overlap.} For the LISA case, we find an overall lensing probability of approximately $0.3\%$ (see Table~\ref{tab:catalog_lensed_event_number_original}), which is slightly smaller than the $\sim 0.5\%$ reported in \cite{2025PhRvD.112l3512G}. We also find that about 60\% of the baseline multiple-image events are reduced to single-image events when the effect of signal overlap is taken into account, indicating a significant impact. For the DECIGO case, we find a lensing probability of approximately $0.15\%$, which is smaller than that in the LISA case. As shown in Table~\ref{tab:decigo_lensed_event_number_original}, the effect of signal overlap is also important for DECIGO. For BNSs and NSBHs, nearly 45\% of the baseline multiple-image events are reduced to single-image events when overlap is considered, whereas for BBHs, only about 4\% of such events are reduced, which is significantly smaller. This difference arises because the signal durations of BBHs in the DECIGO band (see Figure~\ref{Figure:decigo_duration_time_distribution}) are generally much shorter than those of BNSs and NSBHs, leading to a lower probability of signal overlap due to lensing time delays.

\textbf{Statistical results of detected lensed events.} For the LISA case, some formation models originally predict a very small number of intrinsic events (e.g., \texttt{Q3d}). To obtain more robust statistical estimates of the lensed event rates, we therefore randomly sample $10^5$ sets of intrinsic parameters from each model and apply the lensing simulation to these samples. We find that the total number of detected lensed events ranges from 169 to 330 across different models, with the majority producing a single detectable signal. However, a non-negligible fraction of events yield multiple detectable signals due to lensing, with some systems producing up to five detectable signals. The number of events lensed by subhalos is also significant, with approximately 25\% of lensing events involving subhalos, indicating that substructure can substantially affect the observed properties of lensed GWs. In addition, a significant fraction of lensed events have maximum magnification factors greater than 3 ($r_{\mathrm{max}} > 3$), while a smaller subset exhibits extreme magnifications ($r_{\mathrm{max}} > 10$). For the DECIGO case, we find that the total number of detected lensed events ranges from 0 for BNSs to 44 for BBHs. In contrast to the LISA case, some source types in the DECIGO band have shorter signal durations, resulting in a smaller fraction of events affected by signal overlap. In particular, for BBHs, the majority of events produce multiple detectable signals due to lensing, with some systems producing up to five detectable signals. Another difference is that no events exhibit extreme magnifications ($r_{\mathrm{max}} > 10$) in the DECIGO case, which may reflect the different source populations and lensing configurations probed by DECIGO compared to LISA.

\textbf{Limitation on the consideration of the observed signal duration.}
The durations calculated from Eqs.~(\ref{delta_t_obs}) and~(\ref{delta_t_obs_modify}) describe evolution across
the adopted frequency band. A finite observation window can shorten the
portion actually measured and can alter which images contribute to an
overlapping signal. The impact on detectability depends on the omitted
noise-weighted waveform power and on the SNR margin above threshold, rather
than on duration alone. In particular, exceeding an SNR of 8 in Table~\ref{c} is
insufficient to establish insensitivity to this effect.

To quantify the incidence of long-duration signals, we find that
17.12\%, 17.03\%, 26.30\%, 98.28\%, 1.97\%, and 8.93\% of the 10,000 entries
in the public intrinsic catalogs have full-band durations exceeding four
years for HSnodnoSN, HSnodSN, HSnodSNhighaccr, PopIIId, Q3d, and Q3nod,
respectively. Among the systems satisfying our detection criterion in
realization~1 of the public lensed catalogs, the corresponding fractions
are given in Table~\ref{tab:window-sensitivity}. These are unweighted
fractions of catalog entries, not corrected merger rates.

For a simple threshold-sensitivity diagnostic, define
$\rho_{k,\max}=\max_g\rho_{kg}$, where $g$ labels the original signal groups
of system $k$, and $N_0=\sum_k\mathbf{1}(\rho_{k,\max}>8)$.
After an assumed fractional SNR reduction $\epsilon$, the threshold-crossing
fraction is
\begin{equation}
 F_{\mathrm{cross}}(\epsilon)=\frac{1}{N_0}
 \sum_k\mathbf{1}\!\left(8<\rho_{k,\max}\leq
 \frac{8}{1-\epsilon}\right).
 \label{eq:window-stress}
\end{equation}
Here $\mathbf{1}$ denotes the indicator function. All original group
assignments are retained in this diagnostic.
For PopIIId, $F_{\mathrm{cross}}$ is 6.5\%, 14.2\%, and 27.2\% for
$\epsilon=0.05$, 0.10, and 0.20, respectively. In the other five models,
these fractions are zero for reductions up to 10\% and at most 0.3\% for a
20\% reduction in this realization. The PopIIId catalog is particularly
vulnerable: all 169 detectable systems have full-band durations exceeding
four years, 166 have maximum SNRs between 8 and 100, and the median maximum
SNR is 12.47.

The assumed SNR reductions in this diagnostic have not been derived from
an image-dependent observing window. Thus, the threshold-crossing fractions
are conditional sensitivities rather than estimates of actual detection
losses. Full-band duration alone does not determine the SNR reduction, and
these diagnostics cannot be used to rescale the predicted event numbers.
Actual finite-window corrections require the signal power as a function of
frequency, image arrival times, and a consistent treatment of the detector
response and overlapping images.

As a separate timing estimate, suppose that the first geometrical image
coalescence occurs uniformly within a mission of duration $T$. For a system
whose first and last geometrical image coalescences are separated by $D$, the
probability that the last coalescence falls outside the mission is
\begin{equation}
 P_{\mathrm{edge}}(D)=\min\!\left(\frac{D}{T},1\right).
 \label{eq:window-edge}
\end{equation}
Averaging over the detectable systems with multiple geometrical images in
realization~1 gives 10.4\%--14.9\% across the six LISA models. This estimate
includes faint geometrical images and concerns coalescence-time completeness.
It is not a detection-loss fraction, since an image coalescing outside the
mission can still have an observable inspiral.

These results identify long-duration, near-threshold systems, especially
PopIIId, as requiring particular care. The detectability of individual weak
images and the number of separately observable signals can also change when
the system itself remains above threshold. The same finite-window issues
apply to DECIGO; the numerical diagnostics above concern LISA only.
The catalogs therefore provide approximate SNR-selected samples under the
adopted duration prescription. A calculation with a common mission window
for all images is needed to obtain corrected yields and observed signal
multiplicities.

\begin{deluxetable*}{lccccc}
\tablecaption{Duration and conditional threshold-sensitivity diagnostics for realization 1 of the public LISA lensed catalogs.}
\label{tab:window-sensitivity}
\tablehead{
\colhead{Model} &
\colhead{$N_0$} &
\colhead{$N_{\Delta t_{\mathrm{band}}>4\,\mathrm{yr}}/N_0$} &
\colhead{$F_{\mathrm{cross}}(0.05)$} &
\colhead{$F_{\mathrm{cross}}(0.10)$} &
\colhead{$F_{\mathrm{cross}}(0.20)$}
}
\startdata
HSnodnoSN       & 319 & 111/319 (34.8\%) & 0.0\% & 0.0\%  & 0.3\% \\
HSnodSN         & 330 & 113/330 (34.2\%) & 0.0\% & 0.0\%  & 0.0\% \\
HSnodSNhighaccr & 322 & 109/322 (33.9\%) & 0.0\% & 0.0\%  & 0.0\% \\
PopIIId         & 169 & 169/169 (100\%)  & 6.5\% & 14.2\% & 27.2\% \\
Q3d             & 214 & 25/214 (11.7\%)  & 0.0\% & 0.0\%  & 0.0\% \\
Q3nod           & 253 & 46/253 (18.2\%)  & 0.0\% & 0.0\%  & 0.0\%
\enddata
\tablecomments{$N_0$ counts systems with at least one original signal group above SNR 8. The third column uses the catalog full-band inspiral duration. The last three columns give threshold-crossing fractions after assumed uniform SNR reductions, retaining the original signal grouping. These are not actual finite-window detection losses or corrected mission yields. No additional event-rate weighting is applied.}
\end{deluxetable*}

\textbf{Validity of the Geometric-Optics Approximation}. {The transition between the wave- and geometric-optics regimes is primarily controlled by the dimensionless frequency parameter $w$, although the precise boundary can also depend on the lensing configuration. Since the geometric-optics limit follows from the stationary-phase approximation to the diffraction integral, its accuracy may additionally depend on the source position $y$, particularly in the vicinity of caustics\citep{2022JCAP...07..022B}. Nevertheless, this issue is not expected to affect the lens population considered in the present work. In our model, the host halos are restricted to ($M_{\mathrm{hh}}>10^{11}M_{\odot}$), while the subhalos satisfy ($M_{\mathrm{sh}}>10^{10}M_{\odot}$). For the LISA frequency range considered here, $10^{-4}-10^{-1}\,\mathrm{Hz}$, these lens masses correspond to $w\sim10^{2}-10^{9}$, placing the relevant events well within the geometric-optics regime over the parameter space explored in this work. We therefore do not expect wave-optics corrections to appreciably modify our main results. For future observations, however, if the accessible lens population extends toward substantially lower masses or configurations for which diffraction effects become non-negligible, the present framework can be straightforwardly extended by replacing the geometric-optics magnification with the corresponding wave-optics amplification factor.}

By building mock catalogs of strongly lensed GWs for both LISA and DECIGO, we have provided valuable insights into the expected rates and properties of lensed GW events in these future detectors. The main characteristics of our mock catalogs, which could find application in the future work are as follows:
\begin{itemize}
    \item \textbf{Realistic lensing analysis:} As discussed in Section~\ref{sec:lensing_model}, our lensing model incorporates not only galaxy-scale lenses but also cluster-scale halos and subhalo perturbations, providing a more realistic and comprehensive description of lensing effects on GWs. Accordingly, our mock catalogs enable more realistic lensing analyses, including studies of how lensing affects the observed properties of GWs, investigations of the potential of lensed GWs to probe the matter distribution in the Universe, and assessments of the impact of lensing on cosmological parameter estimation.
    
    \item \textbf{Incorporation of signal overlap:} Our analysis demonstrates that signal overlap induced by lensing time delays can significantly affect the observed properties of lensed GWs, particularly for events with long durations and short time delays. By incorporating this effect into our mock catalogs, we provide a more accurate and realistic representation of lensed GW signals, which is essential for developing robust detection and analysis strategies for future observations. In parameter estimation, neglecting signal overlap may lead to misidentifying lensed signals as independent events, thereby introducing biases in the inferred source parameters and population properties. Accounting for signal overlap can therefore improve the accuracy and reliability of parameter estimation for lensed GW events.
    
    \item \textbf{Comprehensive parameter space:} Our mock catalogs span a wide range of intrinsic and extrinsic source parameters, as well as diverse lensing configurations. This comprehensive coverage enables exploration of the full parameter space of lensed GW events and facilitates studies of how source properties, lensing effects, and detector sensitivities jointly influence the observed signals. Such datasets can be directly used, for example, to train machine learning algorithms for identifying lensed GW events or to perform population studies aimed at constraining the properties of GW sources and  distribution of lenses.
\end{itemize}
\section*{Acknowledgments}
This work was supported by National Key R$\&$D Program of China (No. 2024YFC2207400). T.H. Liu was supported  by the National Natural Science Foundation of China under Grants No. 12675061; T.Y. is supported by the National Natural Science Foundation of China Grants No. 12575063. M.B. was supported by the Polish National Science Centre grant 2023/50/A/ST9/00579. We thank Mengfei Sun for useful discussion about response function of DECIGO.

\section*{Data Availability}
Mock catalogs of strongly lensed gravitational wave events generated in this work are publicly available at: \url{https://github.com/LensedGW/GW-LMC-Space}.

\bibliographystyle{aasjournal}
\bibliography{used}

@ARTICLE{2025OJAp....8E...8A,
       author = {{Abe}, Katsuya T. and {Oguri}, Masamune and {Birrer}, Simon and {Khadka}, Narayan and {Marshall}, Philip J. and {Lemon}, Cameron and {More}, Anupreeta and {LSST Dark Energy Science Collaboration}},
        title = "{A halo model approach for mock catalogs of time-variable strong gravitational lenses}",
      journal = {The Open Journal of Astrophysics},
         year = 2025,
        month = jan,
       volume = {8},
          eid = {8},
        pages = {8},
          doi = {10.33232/001c.128482},
archivePrefix = {arXiv},
       eprint = {2411.07509},
 primaryClass = {astro-ph.CO},
       adsurl = {https://ui.adsabs.harvard.edu/abs/2025OJAp....8E...8A}
}

@ARTICLE{OM10,
       author = {{Oguri}, Masamune},
        title = "{Effect of gravitational lensing on the distribution of gravitational waves from distant binary black hole mergers}",
      journal = {\mnras},
         year = 2018,
        month = nov,
       volume = {480},
       number = {3},
        pages = {3842-3855},
          doi = {10.1093/mnras/sty2145},
archivePrefix = {arXiv},
       eprint = {1807.02584},
 primaryClass = {astro-ph.CO},
       adsurl = {https://ui.adsabs.harvard.edu/abs/2018MNRAS.480.3842O}
}

@ARTICLE{2008ApJ...688..709T,
       author = {{Tinker}, Jeremy and {Kravtsov}, Andrey V. and {Klypin}, Anatoly and {Abazajian}, Kevork and {Warren}, Michael and {Yepes}, Gustavo and {Gottl{\"o}ber}, Stefan and {Holz}, Daniel E.},
        title = "{Toward a Halo Mass Function for Precision Cosmology: The Limits of Universality}",
      journal = {\apj},
         year = 2008,
        month = dec,
       volume = {688},
       number = {2},
        pages = {709-728},
          doi = {10.1086/591439},
archivePrefix = {arXiv},
       eprint = {0803.2706},
 primaryClass = {astro-ph},
       adsurl = {https://ui.adsabs.harvard.edu/abs/2008ApJ...688..709T}
}

@ARTICLE{2018ApJS..239...35D,
       author = {{Diemer}, Benedikt},
        title = "{COLOSSUS: A Python Toolkit for Cosmology, Large-scale Structure, and Dark Matter Halos}",
      journal = {\apjs},
         year = 2018,
        month = dec,
       volume = {239},
       number = {2},
          eid = {35},
        pages = {35},
          doi = {10.3847/1538-4365/aaee8c},
archivePrefix = {arXiv},
       eprint = {1712.04512},
 primaryClass = {astro-ph.CO},
       adsurl = {https://ui.adsabs.harvard.edu/abs/2018ApJS..239...35D}
}

@ARTICLE{1996ApJ...462..563N,
       author = {{Navarro}, Julio F. and {Frenk}, Carlos S. and {White}, Simon D.~M.},
        title = "{The Structure of Cold Dark Matter Halos}",
      journal = {\apj},
         year = 1996,
        month = may,
       volume = {462},
        pages = {563},
          doi = {10.1086/177173},
archivePrefix = {arXiv},
       eprint = {astro-ph/9508025},
 primaryClass = {astro-ph},
       adsurl = {https://ui.adsabs.harvard.edu/abs/1996ApJ...462..563N}
}

@ARTICLE{2017NatCo...8.1148L,
       author = {{Liao}, Kai and {Fan}, Xi-Long and {Ding}, Xuheng and {Biesiada}, Marek and {Zhu}, Zong-Hong},
        title = "{Precision cosmology from future lensed gravitational wave and electromagnetic signals}",
      journal = {Nature Communications},
         year = 2017,
        month = oct,
       volume = {8},
          eid = {1148},
        pages = {1148},
          doi = {10.1038/s41467-017-01152-9},
archivePrefix = {arXiv},
       eprint = {1703.04151},
 primaryClass = {astro-ph.CO},
       adsurl = {https://ui.adsabs.harvard.edu/abs/2017NatCo...8.1148L}
}

@article{Maccio2007,
  author  = {Macci{\`o}, A.~V. and Dutton, A.~A. and van den Bosch, F.~C. and
             Moore, B. and Potter, D. and Stadel, J.},
  title   = {Concentration, spin and shape of dark matter haloes: scatter and the dependence on mass and environment},
  journal = {Mon. Not. R. Astron. Soc.},
  volume  = {378},
  pages   = {55--71},
  year    = {2007},
  doi     = {10.1111/j.1365-2966.2007.11720.x}
}

@ARTICLE{2020MNRAS.496.2591O,
       author = {{Okabe}, Taizo and {Oguri}, Masamune and {Peirani}, S{\'e}bastien and {Suto}, Yasushi and {Dubois}, Yohan and {Pichon}, Christophe and {Kitayama}, Tetsu and {Sasaki}, Shin and {Nishimichi}, Takahiro},
        title = "{Shapes and alignments of dark matter haloes and their brightest cluster galaxies in 39 strong lensing clusters}",
      journal = {\mnras},
         year = 2020,
        month = aug,
       volume = {496},
       number = {3},
        pages = {2591-2604},
          doi = {10.1093/mnras/staa1479},
archivePrefix = {arXiv},
       eprint = {2005.11469},
 primaryClass = {astro-ph.CO},
       adsurl = {https://ui.adsabs.harvard.edu/abs/2020MNRAS.496.2591O}
}

@ARTICLE{2013ApJ...762L..31B,
       author = {{Behroozi}, Peter S. and {Wechsler}, Risa H. and {Conroy}, Charlie},
        title = "{On the Lack of Evolution in Galaxy Star Formation Efficiency}",
      journal = {\apjl},
         year = 2013,
        month = jan,
       volume = {762},
       number = {2},
          eid = {L31},
        pages = {L31},
          doi = {10.1088/2041-8205/762/2/L31},
archivePrefix = {arXiv},
       eprint = {1209.3013},
 primaryClass = {astro-ph.CO},
       adsurl = {https://ui.adsabs.harvard.edu/abs/2013ApJ...762L..31B}
}

@ARTICLE{2019MNRAS.488.3143B,
       author = {{Behroozi}, Peter and {Wechsler}, Risa H. and {Hearin}, Andrew P. and {Conroy}, Charlie},
        title = "{UNIVERSEMACHINE: The correlation between galaxy growth and dark matter halo assembly from z = 0-10}",
      journal = {\mnras},
         year = 2019,
        month = sep,
       volume = {488},
       number = {3},
        pages = {3143-3194},
          doi = {10.1093/mnras/stz1182},
archivePrefix = {arXiv},
       eprint = {1806.07893},
 primaryClass = {astro-ph.GA},
       adsurl = {https://ui.adsabs.harvard.edu/abs/2019MNRAS.488.3143B}
}

@ARTICLE{2024ApJ...960...53V,
       author = {{van der Wel}, Arjen and {Martorano}, Marco and {H{\"a}u{\ss}ler}, Boris and {Nedkova}, Kalina V. and {Miller}, Tim B. and {Brammer}, Gabriel B. and {van de Ven}, Glenn and {Leja}, Joel and {Bezanson}, Rachel S. and {Muzzin}, Adam and {Marchesini}, Danilo and {de Graaff}, Anna and {Nelson}, Erica J. and {Kriek}, Mariska and {Bell}, Eric F. and {Franx}, Marijn},
        title = "{Stellar Half-mass Radii of 0.5 z < 2.3 Galaxies: Comparison with JWST/NIRCam Half-light Radii}",
      journal = {\apj},
         year = 2024,
        month = jan,
       volume = {960},
       number = {1},
          eid = {53},
        pages = {53},
          doi = {10.3847/1538-4357/ad02ee},
archivePrefix = {arXiv},
       eprint = {2307.03264},
 primaryClass = {astro-ph.GA},
       adsurl = {https://ui.adsabs.harvard.edu/abs/2024ApJ...960...53V}
}

@ARTICLE{2020ApJ...901...58O,
       author = {{Oguri}, Masamune and {Takahashi}, Ryuichi},
        title = "{Probing Dark Low-mass Halos and Primordial Black Holes with Frequency-dependent Gravitational Lensing Dispersions of Gravitational Waves}",
      journal = {\apj},
         year = 2020,
        month = sep,
       volume = {901},
       number = {1},
          eid = {58},
        pages = {58},
          doi = {10.3847/1538-4357/abafab},
archivePrefix = {arXiv},
       eprint = {2007.01936},
 primaryClass = {astro-ph.CO},
       adsurl = {https://ui.adsabs.harvard.edu/abs/2020ApJ...901...58O}
}

@ARTICLE{2020MNRAS.492.3662I,
       author = {{Ishiyama}, Tomoaki and {Ando}, Shin'ichiro},
        title = "{The abundance and structure of subhaloes near the free streaming scale and their impact on indirect dark matter searches}",
      journal = {\mnras},
         year = 2020,
        month = mar,
       volume = {492},
       number = {3},
        pages = {3662-3671},
          doi = {10.1093/mnras/staa069},
archivePrefix = {arXiv},
       eprint = {1907.03642},
 primaryClass = {astro-ph.CO},
       adsurl = {https://ui.adsabs.harvard.edu/abs/2020MNRAS.492.3662I}
}

@ARTICLE{2015ApJ...799..108D,
       author = {{Diemer}, Benedikt and {Kravtsov}, Andrey V.},
        title = "{A Universal Model for Halo Concentrations}",
      journal = {\apj},
         year = 2015,
        month = jan,
       volume = {799},
       number = {1},
          eid = {108},
        pages = {108},
          doi = {10.1088/0004-637X/799/1/108},
archivePrefix = {arXiv},
       eprint = {1407.4730},
 primaryClass = {astro-ph.CO},
       adsurl = {https://ui.adsabs.harvard.edu/abs/2015ApJ...799..108D}
}

@ARTICLE{2012MNRAS.423.2533B,
       author = {{Barausse}, Enrico},
        title = "{The evolution of massive black holes and their spins in their galactic hosts}",
      journal = {\mnras},
         year = 2012,
        month = jul,
       volume = {423},
       number = {3},
        pages = {2533-2557},
          doi = {10.1111/j.1365-2966.2012.21057.x},
archivePrefix = {arXiv},
       eprint = {1201.5888},
 primaryClass = {astro-ph.CO},
       adsurl = {https://ui.adsabs.harvard.edu/abs/2012MNRAS.423.2533B}
}

@ARTICLE{2025PhRvD.112l3512G,
       author = {{Guti{\'e}rrez}, Juan and {Lagos}, Macarena},
        title = "{Strong-lensing rates of massive black hole binaries in LISA}",
      journal = {\prd},
         year = 2025,
        month = dec,
       volume = {112},
       number = {12},
          eid = {123512},
        pages = {123512},
          doi = {10.1103/yd5h-ql5f},
archivePrefix = {arXiv},
       eprint = {2510.02061},
 primaryClass = {astro-ph.CO},
       adsurl = {https://ui.adsabs.harvard.edu/abs/2025PhRvD.112l3512G}
}

@ARTICLE{2016PhRvD..93b4003K,
       author = {{Klein}, Antoine and {Barausse}, Enrico and {Sesana}, Alberto and {Petiteau}, Antoine and {Berti}, Emanuele and {Babak}, Stanislav and {Gair}, Jonathan and {Aoudia}, Sofiane and {Hinder}, Ian and {Ohme}, Frank and {Wardell}, Barry},
        title = "{Science with the space-based interferometer eLISA: Supermassive black hole binaries}",
      journal = {\prd},
         year = 2016,
        month = jan,
       volume = {93},
       number = {2},
          eid = {024003},
        pages = {024003},
          doi = {10.1103/PhysRevD.93.024003},
archivePrefix = {arXiv},
       eprint = {1511.05581},
 primaryClass = {gr-qc},
       adsurl = {https://ui.adsabs.harvard.edu/abs/2016PhRvD..93b4003K}
}

@ARTICLE{2015ApJ...812...72A,
       author = {{Antonini}, Fabio and {Barausse}, Enrico and {Silk}, Joseph},
        title = "{The Coevolution of Nuclear Star Clusters, Massive Black Holes, and Their Host Galaxies}",
      journal = {\apj},
         year = 2015,
        month = oct,
       volume = {812},
       number = {1},
          eid = {72},
        pages = {72},
          doi = {10.1088/0004-637X/812/1/72},
archivePrefix = {arXiv},
       eprint = {1506.02050},
 primaryClass = {astro-ph.GA},
       adsurl = {https://ui.adsabs.harvard.edu/abs/2015ApJ...812...72A}
}

@ARTICLE{2020ApJ...904...16B,
       author = {{Barausse}, Enrico and {Dvorkin}, Irina and {Tremmel}, Michael and {Volonteri}, Marta and {Bonetti}, Matteo},
        title = "{Massive Black Hole Merger Rates: The Effect of Kiloparsec Separation Wandering and Supernova Feedback}",
      journal = {\apj},
         year = 2020,
        month = nov,
       volume = {904},
       number = {1},
          eid = {16},
        pages = {16},
          doi = {10.3847/1538-4357/abba7f},
archivePrefix = {arXiv},
       eprint = {2006.03065},
 primaryClass = {astro-ph.GA},
       adsurl = {https://ui.adsabs.harvard.edu/abs/2020ApJ...904...16B}
}

@ARTICLE{2023PhRvD.108j3034B,
       author = {{Barausse}, Enrico and {Dey}, Kallol and {Crisostomi}, Marco and {Panayada}, Akshay and {Marsat}, Sylvain and {Basak}, Soumen},
        title = "{Implications of the pulsar timing array detections for massive black hole mergers in the LISA band}",
      journal = {\prd},
         year = 2023,
        month = nov,
       volume = {108},
       number = {10},
          eid = {103034},
        pages = {103034},
          doi = {10.1103/PhysRevD.108.103034},
archivePrefix = {arXiv},
       eprint = {2307.12245},
 primaryClass = {astro-ph.GA},
       adsurl = {https://ui.adsabs.harvard.edu/abs/2023PhRvD.108j3034B}
}

@ARTICLE{1963PhRv..131..435P,
       author = {{Peters}, P.~C. and {Mathews}, J.},
        title = "{Gravitational Radiation from Point Masses in a Keplerian Orbit}",
      journal = {Physical Review},
         year = 1963,
        month = jul,
       volume = {131},
       number = {1},
        pages = {435-440},
          doi = {10.1103/PhysRev.131.435},
       adsurl = {https://ui.adsabs.harvard.edu/abs/1963PhRv..131..435P}
}

@ARTICLE{2003ApJ...595.1039T,
       author = {{Takahashi}, Ryuichi and {Nakamura}, Takashi},
        title = "{Wave Effects in the Gravitational Lensing of Gravitational Waves from Chirping Binaries}",
      journal = {\apj},
         year = 2003,
        month = oct,
       volume = {595},
       number = {2},
        pages = {1039-1051},
          doi = {10.1086/377430},
archivePrefix = {arXiv},
       eprint = {astro-ph/0305055},
 primaryClass = {astro-ph},
       adsurl = {https://ui.adsabs.harvard.edu/abs/2003ApJ...595.1039T}
}

@ARTICLE{2025arXiv251109107S,
       author = {{Sun}, Mengfei and {Wu}, Jie and {Li}, Jin and {Yang}, Nan and {Ma}, Xianghe and {Wang}, Borui and {Zhang}, Minghui and {Zhong}, Yuanhong},
        title = "{Detection of Lensed Gravitational Waves from dark matter halos with deep learning}",
      journal = {arXiv e-prints},
         year = 2025,
        month = nov,
          eid = {arXiv:2511.09107},
        pages = {arXiv:2511.09107},
          doi = {10.48550/arXiv.2511.09107},
archivePrefix = {arXiv},
       eprint = {2511.09107},
 primaryClass = {astro-ph.IM},
       adsurl = {https://ui.adsabs.harvard.edu/abs/2025arXiv251109107S}
}

@ARTICLE{li2026mockcatalogsstronglylensed,
       author = {{Li}, Youkai and {Liao}, Kai and {Sun}, Mingqi and {Yang}, Lilan and {Ding}, Xuheng and {Biesiada}, Marek and {Liu}, Tonghua},
        title = "{Mock Catalogs of Strongly Lensed Gravitational Waves via A Halo Model Approach with Ground-based Detectors}",
      journal = {arXiv e-prints},
         year = 2026,
        month = mar,
          eid = {arXiv:2603.09289},
        pages = {arXiv:2603.09289},
          doi = {10.48550/arXiv.2603.09289},
archivePrefix = {arXiv},
       eprint = {2603.09289},
 primaryClass = {astro-ph.CO},
       adsurl = {https://ui.adsabs.harvard.edu/abs/2026arXiv260309289L}
}

@ARTICLE{2013ApJ...779...72D,
       author = {{Dominik}, Michal and {Belczynski}, Krzysztof and {Fryer}, Christopher and {Holz}, Daniel E. and {Berti}, Emanuele and {Bulik}, Tomasz and {Mandel}, Ilya and {O'Shaughnessy}, Richard},
        title = "{Double Compact Objects. II. Cosmological Merger Rates}",
      journal = {\apj},
         year = 2013,
        month = dec,
       volume = {779},
       number = {1},
          eid = {72},
        pages = {72},
          doi = {10.1088/0004-637X/779/1/72},
archivePrefix = {arXiv},
       eprint = {1308.1546},
 primaryClass = {astro-ph.HE},
       adsurl = {https://ui.adsabs.harvard.edu/abs/2013ApJ...779...72D},
}

@ARTICLE{2011PhRvD..83d4011Y,
       author = {{Yagi}, Kent and {Seto}, Naoki},
        title = "{Detector configuration of DECIGO/BBO and identification of cosmological neutron-star binaries}",
      journal = {\prd},
         year = 2011,
        month = feb,
       volume = {83},
       number = {4},
          eid = {044011},
        pages = {044011},
          doi = {10.1103/PhysRevD.83.044011},
archivePrefix = {arXiv},
       eprint = {1101.3940},
 primaryClass = {astro-ph.CO},
       adsurl = {https://ui.adsabs.harvard.edu/abs/2011PhRvD..83d4011Y}
}

@ARTICLE{2021Galax...9...14I,
       author = {{Ishikawa}, Tomohiro and {Iwaguchi}, Shoki and {Michimura}, Yuta and {Ando}, Masaki and {Yamada}, Rika and {Watanabe}, Izumi and {Nagano}, Koji and {Akutsu}, Tomotada and {Komori}, Kentaro and {Musha}, Mitsuru and {Naito}, Takeo and {Morimoto}, Taigen and {Kawamura}, Seiji},
        title = "{Improvement of the Target Sensitivity in DECIGO by Optimizing Its Parameters for Quantum Noise Including the Effect of Diffraction Loss}",
      journal = {Galaxies},
         year = 2021,
        month = feb,
       volume = {9},
       number = {1},
          eid = {14},
        pages = {14},
          doi = {10.3390/galaxies9010014},
archivePrefix = {arXiv},
       eprint = {2012.11859},
 primaryClass = {gr-qc},
       adsurl = {https://ui.adsabs.harvard.edu/abs/2021Galax...9...14I}
}

@ARTICLE{2016PhRvL.116f1102A,
       author = {{Abbott}, B.~P. and {Abbott}, R. and {Abbott}, T.~D. and {Abernathy}, M.~R. and {Acernese}, F. and {Ackley}, K. and {Adams}, C. and {Adams}, T. and {Addesso}, P. and {Adhikari}, R.~X. and {Adya}, V.~B. and {Affeldt}, C. and {Agathos}, M. and {Agatsuma}, K. and {Aggarwal}, N. and {Aguiar}, O.~D. and {Aiello}, L. and {Ain}, A. and {Ajith}, P. and {Allen}, B. and {Allocca}, A. and {Altin}, P.~A. and {Anderson}, S.~B. and {Anderson}, W.~G. and {Arai}, K. and {Arain}, M.~A. and {Araya}, M.~C. and {Arceneaux}, C.~C. and {Areeda}, J.~S. and {Arnaud}, N. and {Arun}, K.~G. and {Ascenzi}, S. and {Ashton}, G. and {Ast}, M. and {Aston}, S.~M. and {Astone}, P. and {Aufmuth}, P. and {Aulbert}, C. and {Babak}, S. and {Bacon}, P. and {Bader}, M.~K.~M. and {Baker}, P.~T. and {Baldaccini}, F. and {Ballardin}, G. and {Ballmer}, S.~W. and {Barayoga}, J.~C. and {Barclay}, S.~E. and {Barish}, B.~C. and {Barker}, D. and {Barone}, F. and {Barr}, B. and {Barsotti}, L. and {Barsuglia}, M. and {Barta}, D. and {Bartlett}, J. and {Barton}, M.~A. and {Bartos}, I. and {Bassiri}, R. and {Basti}, A. and {Batch}, J.~C. and {Baune}, C. and {Bavigadda}, V. and {Bazzan}, M. and {Behnke}, B. and {Bejger}, M. and {Belczynski}, C. and {Bell}, A.~S. and {Bell}, C.~J. and {Berger}, B.~K. and {Bergman}, J. and {Bergmann}, G. and {Berry}, C.~P.~L. and {Bersanetti}, D. and {Bertolini}, A. and {Betzwieser}, J. and {Bhagwat}, S. and {Bhandare}, R. and {Bilenko}, I.~A. and {Billingsley}, G. and {Birch}, J. and {Birney}, I.~A. and {Birnholtz}, O. and {Biscans}, S. and {Bisht}, A. and {Bitossi}, M. and {Biwer}, C. and {Bizouard}, M.~A. and {Blackburn}, J.~K. and {Blair}, C.~D. and {Blair}, D.~G. and {Blair}, R.~M. and {Bloemen}, S. and {Bock}, O. and {Bodiya}, T.~P. and {Boer}, M. and {Bogaert}, G. and {Bogan}, C. and {Bohe}, A. and {Bojtos}, P. and {Bond}, C. and {Bondu}, F. and {Bonnand}, R. and {Boom}, B.~A. and {Bork}, R. and {Boschi}, V. and {Bose}, S. and {Bouffanais}, Y. and {Bozzi}, A. and {Bradaschia}, C. and {Brady}, P.~R. and {Braginsky}, V.~B. and {Branchesi}, M. and {Brau}, J.~E. and {Briant}, T. and {Brillet}, A. and {Brinkmann}, M. and {Brisson}, V. and {Brockill}, P. and {Brooks}, A.~F. and {Brown}, D.~A. and {Brown}, D.~D. and {Brown}, N.~M. and {Buchanan}, C.~C. and {Buikema}, A. and {Bulik}, T. and {Bulten}, H.~J. and {Buonanno}, A. and {Buskulic}, D. and {Buy}, C. and {Byer}, R.~L. and {Cabero}, M. and {Cadonati}, L. and {Cagnoli}, G. and {Cahillane}, C. and {Bustillo}, J. Calder{\'o}n and {Callister}, T. and {Calloni}, E. and {Camp}, J.~B. and {Cannon}, K.~C. and {Cao}, J. and {Capano}, C.~D. and {Capocasa}, E. and {Carbognani}, F. and {Caride}, S. and {Diaz}, J. Casanueva and {Casentini}, C. and {Caudill}, S. and {Cavagli{\`a}}, M. and {Cavalier}, F. and {Cavalieri}, R. and {Cella}, G. and {Cepeda}, C.~B. and {Baiardi}, L. Cerboni and {Cerretani}, G. and {Cesarini}, E. and {Chakraborty}, R. and {Chalermsongsak}, T. and {Chamberlin}, S.~J. and {Chan}, M. and {Chao}, S. and {Charlton}, P. and {Chassande-Mottin}, E. and {Chen}, H.~Y. and {Chen}, Y. and {Cheng}, C. and {Chincarini}, A. and {Chiummo}, A. and {Cho}, H.~S. and {Cho}, M. and {Chow}, J.~H. and {Christensen}, N. and {Chu}, Q. and {Chua}, S. and {Chung}, S. and {Ciani}, G. and {Clara}, F. and {Clark}, J.~A. and {Cleva}, F. and {Coccia}, E. and {Cohadon}, P.-F. and {Colla}, A. and {Collette}, C.~G. and {Cominsky}, L. and {Constancio}, M. and {Conte}, A. and {Conti}, L. and {Cook}, D. and {Corbitt}, T.~R. and {Cornish}, N. and {Corsi}, A. and {Cortese}, S. and {Costa}, C.~A. and {Coughlin}, M.~W. and {Coughlin}, S.~B. and {Coulon}, J.-P. and {Countryman}, S.~T. and {Couvares}, P. and {Cowan}, E.~E. and {Coward}, D.~M. and {Cowart}, M.~J.},
        title = "{Observation of Gravitational Waves from a Binary Black Hole Merger}",
      journal = {\prl},
         year = 2016,
        month = feb,
       volume = {116},
       number = {6},
          eid = {061102},
        pages = {061102},
          doi = {10.1103/PhysRevLett.116.061102},
archivePrefix = {arXiv},
       eprint = {1602.03837},
 primaryClass = {gr-qc},
       adsurl = {https://ui.adsabs.harvard.edu/abs/2016PhRvL.116f1102A}
}

@ARTICLE{2025arXiv250818082T,
       author = {{The LIGO Scientific Collaboration} and {the Virgo Collaboration} and {the KAGRA Collaboration} and {Abac}, A.~G. and {Abouelfettouh}, I. and {Acernese}, F. and {Ackley}, K. and {Adamcewicz}, C. and {Adhicary}, S. and {Adhikari}, D. and {Adhikari}, N. and {Adhikari}, R.~X. and {Adkins}, V.~K. and {Afroz}, S. and {Agapito}, A. and {Agarwal}, D. and {Agathos}, M. and {Aggarwal}, N. and {Aggarwal}, S. and {Aguiar}, O.~D. and {Ahrend}, I.-L. and {Aiello}, L. and {Ain}, A. and {Ajith}, P. and {Akutsu}, T. and {Albanesi}, S. and {Ali}, W. and {Al-Kershi}, S. and {All{\'e}n{\'e}}, C. and {Allocca}, A. and {Al-Shammari}, S. and {Altin}, P.~A. and {Alvarez-Lopez}, S. and {Amar}, W. and {Amarasinghe}, O. and {Amato}, A. and {Amicucci}, F. and {Amra}, C. and {Ananyeva}, A. and {Anderson}, S.~B. and {Anderson}, W.~G. and {Andia}, M. and {Ando}, M. and {Andr{\'e}s-Carcasona}, M. and {Andri{\'c}}, T. and {Anglin}, J. and {Ansoldi}, S. and {Antelis}, J.~M. and {Antier}, S. and {Aoumi}, M. and {Appavuravther}, E.~Z. and {Appert}, S. and {Apple}, S.~K. and {Arai}, K. and {Araya}, A. and {Araya}, M.~C. and {Arca Sedda}, M. and {Areeda}, J.~S. and {Aritomi}, N. and {Armato}, F. and {Armstrong}, S. and {Arnaud}, N. and {Arogeti}, M. and {Aronson}, S.~M. and {Arun}, K.~G. and {Ashton}, G. and {Aso}, Y. and {Asprea}, L. and {Assiduo}, M. and {Assis de Souza Melo}, S. and {Aston}, S.~M. and {Astone}, P. and {Attadio}, F. and {Aubin}, F. and {AultONeal}, K. and {Avallone}, G. and {Avila}, E.~A. and {Babak}, S. and {Badger}, C. and {Bae}, S. and {Bagnasco}, S. and {Baiotti}, L. and {Bajpai}, R. and {Baka}, T. and {Baker}, A.~M. and {Baker}, K.~A. and {Baker}, T. and {Baldi}, G. and {Baldicchi}, N. and {Ball}, M. and {Ballardin}, G. and {Ballmer}, S.~W. and {Banagiri}, S. and {Banerjee}, B. and {Bankar}, D. and {Baptiste}, T.~M. and {Baral}, P. and {Baratti}, M. and {Barayoga}, J.~C. and {Barish}, B.~C. and {Barker}, D. and {Barman}, N. and {Barneo}, P. and {Barone}, F. and {Barr}, B. and {Barsotti}, L. and {Barsuglia}, M. and {Barta}, D. and {Bartoletti}, A.~M. and {Barton}, M.~A. and {Bartos}, I. and {Basalaev}, A. and {Bassiri}, R. and {Basti}, A. and {Bawaj}, M. and {Baxi}, P. and {Bayley}, J.~C. and {Baylor}, A.~C. and {Baynard}, II, P.~A. and {Bazzan}, M. and {Bedakihale}, V.~M. and {Beirnaert}, F. and {Bejger}, M. and {Belardinelli}, D. and {Bell}, A.~S. and {Bellie}, D.~S. and {Bellizzi}, L. and {Benoit}, W. and {Bentara}, I. and {Bentley}, J.~D. and {Ben Yaala}, M. and {Bera}, S. and {Bergamin}, F. and {Berger}, B.~K. and {Bernuzzi}, S. and {Beroiz}, M. and {Berry}, C.~P.~L. and {Bersanetti}, D. and {Bertheas}, T. and {Bertolini}, A. and {Betzwieser}, J. and {Beveridge}, D. and {Bevilacqua}, G. and {Bevins}, N. and {Bhandare}, R. and {Bhatt}, R. and {Bhattacharjee}, D. and {Bhattacharyya}, S. and {Bhaumik}, S. and {Biancalana}, V. and {Bianchi}, A. and {Bilenko}, I.~A. and {Billingsley}, G. and {Binetti}, A. and {Bini}, S. and {Binu}, C. and {Biot}, S. and {Birnholtz}, O. and {Biscoveanu}, S. and {Bisht}, A. and {Bitossi}, M. and {Bizouard}, M.-A. and {Blaber}, S. and {Blackburn}, J.~K. and {Blagg}, L.~A. and {Blair}, C.~D. and {Blair}, D.~G. and {Bode}, N. and {Boettner}, N. and {Boileau}, G. and {Boldrini}, M. and {Bolingbroke}, G.~N. and {Bolliand}, A. and {Bonavena}, L.~D. and {Bondarescu}, R. and {Bondu}, F. and {Bonilla}, E. and {Bonilla}, M.~S. and {Bonino}, A. and {Bonnand}, R. and {Borchers}, A. and {Borhanian}, S. and {Boschi}, V. and {Bose}, S. and {Bossilkov}, V. and {Bothra}, Y. and {Boudon}, A. and {Bourg}, L. and {Boyle}, M. and {Bozzi}, A. and {Bradaschia}, C. and {Brady}, P.~R. and {Branch}, A. and {Branchesi}, M. and {Braun}, I. and {Briant}, T. and {Brillet}, A. and {Brinkmann}, M. and {Brockill}, P. and {Brockmueller}, E.},
        title = "{GWTC-4.0: Updating the Gravitational-Wave Transient Catalog with Observations from the First Part of the Fourth LIGO-Virgo-KAGRA Observing Run}",
      journal = {arXiv e-prints},
         year = 2025,
        month = aug,
          eid = {arXiv:2508.18082},
        pages = {arXiv:2508.18082},
          doi = {10.48550/arXiv.2508.18082},
archivePrefix = {arXiv},
       eprint = {2508.18082},
 primaryClass = {gr-qc},
       adsurl = {https://ui.adsabs.harvard.edu/abs/2025arXiv250818082T}
}

@ARTICLE{2025arXiv251216347T,
       author = {{The LIGO Scientific Collaboration} and {the Virgo Collaboration} and {the KAGRA Collaboration} and {Abac}, A.~G. and {Abouelfettouh}, I. and {Acernese}, F. and {Ackley}, K. and {Adamcewicz}, C. and {Adhicary}, S. and {Adhikari}, D. and {Adhikari}, N. and {Adhikari}, R.~X. and {Adkins}, V.~K. and {Afroz}, S. and {Agapito}, A. and {Agarwal}, D. and {Agathos}, M. and {Aggarwal}, N. and {Aggarwal}, S. and {Aguiar}, O.~D. and {Ahrend}, I.-L. and {Aiello}, L. and {Ain}, A. and {Ajith}, P. and {Akutsu}, T. and {Albanesi}, S. and {Ali}, W. and {Al-Kershi}, S. and {All{\'e}n{\'e}}, C. and {Allocca}, A. and {Al-Shammari}, S. and {Altin}, P.~A. and {Alvarez-Lopez}, S. and {Amar}, W. and {Amarasinghe}, O. and {Amato}, A. and {Amicucci}, F. and {Amra}, C. and {Ananyeva}, A. and {Anderson}, S.~B. and {Anderson}, W.~G. and {Andia}, M. and {Ando}, M. and {Andr{\'e}s-Carcasona}, M. and {Andri{\'c}}, T. and {Anglin}, J. and {Ansoldi}, S. and {Antelis}, J.~M. and {Antier}, S. and {Aoumi}, M. and {Appavuravther}, E.~Z. and {Appert}, S. and {Apple}, S.~K. and {Arai}, K. and {Araya}, A. and {Araya}, M.~C. and {Arca Sedda}, M. and {Areeda}, J.~S. and {Aritomi}, N. and {Armato}, F. and {Armstrong}, S. and {Arnaud}, N. and {Arogeti}, M. and {Aronson}, S.~M. and {Ashton}, G. and {Aso}, Y. and {Asprea}, L. and {Assiduo}, M. and {Assis de Souza Melo}, S. and {Aston}, S.~M. and {Astone}, P. and {Attadio}, F. and {Aubin}, F. and {AultONeal}, K. and {Avallone}, G. and {Avila}, E.~A. and {Babak}, S. and {Badger}, C. and {Bae}, S. and {Bagnasco}, S. and {Baiotti}, L. and {Bajpai}, R. and {Baka}, T. and {Baker}, A.~M. and {Baker}, K.~A. and {Baker}, T. and {Baldi}, G. and {Baldicchi}, N. and {Ball}, M. and {Ballardin}, G. and {Ballmer}, S.~W. and {Banagiri}, S. and {Banerjee}, B. and {Bankar}, D. and {Baptiste}, T.~M. and {Baral}, P. and {Baratti}, M. and {Barayoga}, J.~C. and {Barish}, B.~C. and {Barker}, D. and {Barman}, N. and {Barneo}, P. and {Barone}, F. and {Barr}, B. and {Barsode}, A. and {Barsotti}, L. and {Barsuglia}, M. and {Barta}, D. and {Bartoletti}, A.~M. and {Barton}, M.~A. and {Bartos}, I. and {Basak}, S. and {Basalaev}, A. and {Bassiri}, R. and {Basti}, A. and {Bawaj}, M. and {Baxi}, P. and {Bayley}, J.~C. and {Baylor}, A.~C. and {Baynard}, II, P.~A. and {Bazzan}, M. and {Bedakihale}, V.~M. and {Beirnaert}, F. and {Bejger}, M. and {Belardinelli}, D. and {Bell}, A.~S. and {Bellie}, D.~S. and {Bellizzi}, L. and {Benoit}, W. and {Bentara}, I. and {Bentley}, J.~D. and {Ben Yaala}, M. and {Bera}, S. and {Bergamin}, F. and {Berger}, B.~K. and {Bernuzzi}, S. and {Beroiz}, M. and {Berry}, C.~P.~L. and {Bersanetti}, D. and {Bertheas}, T. and {Bertolini}, A. and {Betzwieser}, J. and {Beveridge}, D. and {Bevilacqua}, G. and {Bevins}, N. and {Bhandare}, R. and {Bhatt}, R. and {Bhattacharjee}, D. and {Bhattacharyya}, S. and {Bhaumik}, S. and {Biancalana}, V. and {Bianchi}, A. and {Bilenko}, I.~A. and {Billingsley}, G. and {Binetti}, A. and {Binu}, C. and {Biot}, S. and {Birnholtz}, O. and {Biscoveanu}, S. and {Bisht}, A. and {Bitossi}, M. and {Bizouard}, M.-A. and {Blaber}, S. and {Blackburn}, J.~K. and {Blagg}, L.~A. and {Blair}, C.~D. and {Blair}, D.~G. and {Bode}, N. and {Boettner}, N. and {Boileau}, G. and {Boldrini}, M. and {Bolingbroke}, G.~N. and {Bolliand}, A. and {Bonavena}, L.~D. and {Bondarescu}, R. and {Bondu}, F. and {Bonilla}, E. and {Bonilla}, M.~S. and {Bonino}, A. and {Bonnand}, R. and {Borchers}, A. and {Borhanian}, S. and {Boschi}, V. and {Bose}, S. and {Bossilkov}, V. and {Bothra}, Y. and {Boudon}, A. and {Bourg}, L. and {Boyle}, M. and {Bozzi}, A. and {Bradaschia}, C. and {Brady}, P.~R. and {Branch}, A. and {Branchesi}, M. and {Braun}, I. and {Briant}, T. and {Brillet}, A. and {Brinkmann}, M. and {Brockill}, P. and {Brockmueller}, E.},
        title = "{GWTC-4.0: Searches for Gravitational-Wave Lensing Signatures}",
      journal = {arXiv e-prints},
         year = 2025,
        month = dec,
          eid = {arXiv:2512.16347},
        pages = {arXiv:2512.16347},
          doi = {10.48550/arXiv.2512.16347},
archivePrefix = {arXiv},
       eprint = {2512.16347},
 primaryClass = {gr-qc},
       adsurl = {https://ui.adsabs.harvard.edu/abs/2025arXiv251216347T}
}

@ARTICLE{2025arXiv250904348T,
       author = {{The LIGO Scientific Collaboration} and {the Virgo Collaboration} and {the KAGRA Collaboration} and {Abac}, A.~G. and {Abouelfettouh}, I. and {Acernese}, F. and {Ackley}, K. and {Adamcewicz}, C. and {Adhicary}, S. and {Adhikari}, D. and {Adhikari}, N. and {Adhikari}, R.~X. and {Adkins}, V.~K. and {Afroz}, S. and {Agapito}, A. and {Agarwal}, D. and {Agathos}, M. and {Aggarwal}, N. and {Aggarwal}, S. and {Aguiar}, O.~D. and {Ahrend}, I.-L. and {Aiello}, L. and {Ain}, A. and {Ajith}, P. and {Akutsu}, T. and {Albanesi}, S. and {Ali}, W. and {Al-Kershi}, S. and {All{\'e}n{\'e}}, C. and {Allocca}, A. and {Al-Shammari}, S. and {Altin}, P.~A. and {Alvarez-Lopez}, S. and {Amar}, W. and {Amarasinghe}, O. and {Amato}, A. and {Amicucci}, F. and {Amra}, C. and {Ananyeva}, A. and {Anderson}, S.~B. and {Anderson}, W.~G. and {Andia}, M. and {Ando}, M. and {Andr{\'e}s-Carcasona}, M. and {Andri{\'c}}, T. and {Anglin}, J. and {Ansoldi}, S. and {Antelis}, J.~M. and {Antier}, S. and {Aoumi}, M. and {Appavuravther}, E.~Z. and {Appert}, S. and {Apple}, S.~K. and {Arai}, K. and {Araya}, A. and {Araya}, M.~C. and {Arca Sedda}, M. and {Areeda}, J.~S. and {Aritomi}, N. and {Armato}, F. and {Armstrong}, S. and {Arnaud}, N. and {Arogeti}, M. and {Aronson}, S.~M. and {Arun}, K.~G. and {Ashton}, G. and {Aso}, Y. and {Asprea}, L. and {Assiduo}, M. and {Assis de Souza Melo}, S. and {Aston}, S.~M. and {Astone}, P. and {Attadio}, F. and {Aubin}, F. and {AultONeal}, K. and {Avallone}, G. and {Avila}, E.~A. and {Babak}, S. and {Badger}, C. and {Bae}, S. and {Bagnasco}, S. and {Baiotti}, L. and {Bajpai}, R. and {Baka}, T. and {Baker}, A.~M. and {Baker}, K.~A. and {Baker}, T. and {Baldi}, G. and {Baldicchi}, N. and {Ball}, M. and {Ballardin}, G. and {Ballmer}, S.~W. and {Banagiri}, S. and {Banerjee}, B. and {Bankar}, D. and {Baptiste}, T.~M. and {Baral}, P. and {Baratti}, M. and {Barayoga}, J.~C. and {Barish}, B.~C. and {Barker}, D. and {Barman}, N. and {Barneo}, P. and {Barone}, F. and {Barr}, B. and {Barsotti}, L. and {Barsuglia}, M. and {Barta}, D. and {Bartoletti}, A.~M. and {Barton}, M.~A. and {Bartos}, I. and {Basalaev}, A. and {Bassiri}, R. and {Basti}, A. and {Bawaj}, M. and {Baxi}, P. and {Bayley}, J.~C. and {Baylor}, A.~C. and {Baynard}, II, P.~A. and {Bazzan}, M. and {Bedakihale}, V.~M. and {Beirnaert}, F. and {Bejger}, M. and {Belardinelli}, D. and {Bell}, A.~S. and {Bellie}, D.~S. and {Bellizzi}, L. and {Benoit}, W. and {Bentara}, I. and {Bentley}, J.~D. and {Ben Yaala}, M. and {Bera}, S. and {Bergamin}, F. and {Berger}, B.~K. and {Bernuzzi}, S. and {Beroiz}, M. and {Berry}, C.~P.~L. and {Bersanetti}, D. and {Bertheas}, T. and {Bertolini}, A. and {Betzwieser}, J. and {Beveridge}, D. and {Bevilacqua}, G. and {Bevins}, N. and {Bhandare}, R. and {Bhatt}, R. and {Bhattacharjee}, D. and {Bhattacharyya}, S. and {Bhaumik}, S. and {Biancalana}, V. and {Bianchi}, A. and {Bilenko}, I.~A. and {Bilicki}, M. and {Billingsley}, G. and {Binetti}, A. and {Bini}, S. and {Binu}, C. and {Biot}, S. and {Birnholtz}, O. and {Biscoveanu}, S. and {Bisht}, A. and {Bitossi}, M. and {Bizouard}, M.-A. and {Blaber}, S. and {Blackburn}, J.~K. and {Blagg}, L.~A. and {Blair}, C.~D. and {Blair}, D.~G. and {Bode}, N. and {Boettner}, N. and {Boileau}, G. and {Boldrini}, M. and {Bolingbroke}, G.~N. and {Bolliand}, A. and {Bonavena}, L.~D. and {Bondarescu}, R. and {Bondu}, F. and {Bonilla}, E. and {Bonilla}, M.~S. and {Bonino}, A. and {Bonnand}, R. and {Borchers}, A. and {Borhanian}, S. and {Boschi}, V. and {Bose}, S. and {Bossilkov}, V. and {Bothra}, Y. and {Boudon}, A. and {Bourg}, L. and {Boyle}, M. and {Bozzi}, A. and {Bradaschia}, C. and {Brady}, P.~R. and {Branch}, A. and {Branchesi}, M. and {Braun}, I. and {Briant}, T. and {Brillet}, A. and {Brinkmann}, M. and {Brockill}, P.},
        title = "{GWTC-4.0: Constraints on the Cosmic Expansion Rate and Modified Gravitational-wave Propagation}",
      journal = {arXiv e-prints},
         year = 2025,
        month = sep,
          eid = {arXiv:2509.04348},
        pages = {arXiv:2509.04348},
          doi = {10.48550/arXiv.2509.04348},
archivePrefix = {arXiv},
       eprint = {2509.04348},
 primaryClass = {astro-ph.CO},
       adsurl = {https://ui.adsabs.harvard.edu/abs/2025arXiv250904348T}
}

@ARTICLE{2020GReGr..52...81B,
       author = {{Barausse}, Enrico and {Berti}, Emanuele and {Hertog}, Thomas and {Hughes}, Scott A. and {Jetzer}, Philippe and {Pani}, Paolo and {Sotiriou}, Thomas P. and {Tamanini}, Nicola and {Witek}, Helvi and {Yagi}, Kent and {Yunes}, Nicol{\'a}s and {Abdelsalhin}, T. and {Achucarro}, A. and {van Aelst}, K. and {Afshordi}, N. and {Akcay}, S. and {Annulli}, L. and {Arun}, K.~G. and {Ayuso}, I. and {Baibhav}, V. and {Baker}, T. and {Bantilan}, H. and {Barreiro}, T. and {Barrera-Hinojosa}, C. and {Bartolo}, N. and {Baumann}, D. and {Belgacem}, E. and {Bellini}, E. and {Bellomo}, N. and {Ben-Dayan}, I. and {Bena}, I. and {Benkel}, R. and {Bergshoefs}, E. and {Bernard}, L. and {Bernuzzi}, S. and {Bertacca}, D. and {Besancon}, M. and {Beutler}, F. and {Beyer}, F. and {Bhagwat}, S. and {Bicak}, J. and {Biondini}, S. and {Bize}, S. and {Blas}, D. and {Boehmer}, C. and {Boller}, K. and {Bonga}, B. and {Bonvin}, C. and {Bosso}, P. and {Bozzola}, G. and {Brax}, P. and {Breitbach}, M. and {Brito}, R. and {Bruni}, M. and {Br{\"u}gmann}, B. and {Bulten}, H. and {Buonanno}, A. and {Burko}, L.~M. and {Burrage}, C. and {Cabral}, F. and {Calcagni}, G. and {Caprini}, C. and {C{\'a}rdenas-Avenda{\~n}o}, A. and {Celoria}, M. and {Chatziioannou}, K. and {Chernoff}, D. and {Clough}, K. and {Coates}, A. and {Comelli}, D. and {Comp{\`e}re}, G. and {Croon}, D. and {Cruces}, D. and {Cusin}, G. and {Dalang}, C. and {Danielsson}, U. and {Das}, S. and {Datta}, S. and {de Boer}, J. and {De Luca}, V. and {De Rham}, C. and {Desjacques}, V. and {Destounis}, K. and {Di Filippo}, F. and {Dima}, A. and {Dimastrogiovanni}, E. and {Dolan}, S. and {Doneva}, D. and {Duque}, F. and {Durrer}, R. and {East}, W. and {Easther}, R. and {Elley}, M. and {Ellis}, J.~R. and {Emparan}, R. and {Ezquiaga}, J.~M. and {Fairbairn}, M. and {Fairhurst}, S. and {Farmer}, H.~F. and {Fasiello}, M.~R. and {Ferrari}, V. and {Ferreira}, P.~G. and {Ficarra}, G. and {Figueras}, P. and {Fisenko}, S. and {Foffa}, S. and {Franchini}, N. and {Franciolini}, G. and {Fransen}, K. and {Frauendiener}, J. and {Frusciante}, N. and {Fujita}, R. and {Gair}, J. and {Ganz}, A. and {Garcia}, P. and {Garcia-Bellido}, J. and {Garriga}, J. and {Geiger}, R. and {Geng}, C. and {Gergely}, L. {\'A}. and {Germani}, C. and {Gerosa}, D. and {Giddings}, S.~B. and {Gourgoulhon}, E. and {Grandclement}, P. and {Graziani}, L. and {Gualtieri}, L. and {Haggard}, D. and {Haino}, S. and {Halburd}, R. and {Han}, W.-B. and {Hawken}, A.~J. and {Hees}, A. and {Heng}, I.~S. and {Hennig}, J. and {Herdeiro}, C. and {Hervik}, S. and {Holten}, J. v. and {Hoyle}, C.~D. and {Hu}, Y. and {Hull}, M. and {Ikeda}, T. and {Isi}, M. and {Jenkins}, A. and {Juli{\'e}}, F. and {Kajfasz}, E. and {Kalaghatgi}, C. and {Kaloper}, N. and {Kamionkowski}, M. and {Karas}, V. and {Kastha}, S. and {Keresztes}, Z. and {Kidder}, L. and {Kimpson}, T. and {Klein}, A. and {Klioner}, S. and {Kokkotas}, K. and {Kolesova}, H. and {Kolkowitz}, S. and {Kopp}, J. and {Koyama}, K. and {Krishnendu}, N.~V. and {Kroon}, J.~A.~V. and {Kunz}, M. and {Lahav}, O. and {Landragin}, A. and {Lang}, R.~N. and {Le Poncin-Lafitte}, C. and {Lemos}, J. and {Li}, B. and {Liberati}, S. and {Liguori}, M. and {Lin}, F. and {Liu}, G. and {Lobo}, F.~S.~N. and {Loll}, R. and {Lombriser}, L. and {Lovelace}, G. and {Macedo}, R.~P. and {Madge}, E. and {Maggio}, E. and {Maggiore}, M. and {Marassi}, S. and {Marcoccia}, P. and {Markakis}, C. and {Martens}, W. and {Martinovic}, K. and {Martins}, C.~J.~A.~P. and {Maselli}, A. and {Mastrogiovanni}, S. and {Matarrese}, S. and {Matas}, A. and {Mavromatos}, N.~E. and {Mazumdar}, A. and {Meerburg}, P.~D. and {Megias}, E. and {Miller}, J. and {Mimoso}, J.~P. and {Mittnacht}, L. and {Montero}, M.~M. and {Moore}, B.},
        title = "{Prospects for fundamental physics with LISA}",
      journal = {General Relativity and Gravitation},
         year = 2020,
        month = aug,
       volume = {52},
       number = {8},
          eid = {81},
        pages = {81},
          doi = {10.1007/s10714-020-02691-1},
archivePrefix = {arXiv},
       eprint = {2001.09793},
 primaryClass = {gr-qc},
       adsurl = {https://ui.adsabs.harvard.edu/abs/2020GReGr..52...81B}
}

@ARTICLE{2019PhRvD.100h4002H,
       author = {{Hohmann}, Manuel and {J{\"a}rv}, Laur and {Kr{\v{s}}{\v{s}}{\'a}k}, Martin and {Pfeifer}, Christian},
        title = "{Modified teleparallel theories of gravity in symmetric spacetimes}",
      journal = {\prd},
         year = 2019,
        month = oct,
       volume = {100},
       number = {8},
          eid = {084002},
        pages = {084002},
          doi = {10.1103/PhysRevD.100.084002},
archivePrefix = {arXiv},
       eprint = {1901.05472},
 primaryClass = {gr-qc},
       adsurl = {https://ui.adsabs.harvard.edu/abs/2019PhRvD.100h4002H}
}

@ARTICLE{2026PhRvL.136j1003C,
       author = {{Cousins}, Bryce and {Schumacher}, Kristen and {Chung}, Adrian Ka-Wai and {Talbot}, Colm and {Callister}, Thomas and {Holz}, Daniel E. and {Yunes}, Nicol{\'a}s},
        title = "{Stochastic Siren: Astrophysical Gravitational-Wave Background Measurements of the Hubble Constant}",
      journal = {\prl},
         year = 2026,
        month = mar,
       volume = {136},
       number = {10},
          eid = {101003},
        pages = {101003},
          doi = {10.1103/4lzh-bm7y},
archivePrefix = {arXiv},
       eprint = {2503.01997},
 primaryClass = {astro-ph.CO},
       adsurl = {https://ui.adsabs.harvard.edu/abs/2026PhRvL.136j1003C}
}

@ARTICLE{2025PhRvD.112b3002S,
       author = {{Sun}, Mingqi and {Liao}, Kai},
        title = "{Probing stellar-mass primordial black holes with type Ia supernova microlensing}",
      journal = {\prd},
         year = 2025,
        month = jul,
       volume = {112},
       number = {2},
          eid = {023002},
        pages = {023002},
          doi = {10.1103/xll3-68j2},
archivePrefix = {arXiv},
       eprint = {2503.22341},
 primaryClass = {astro-ph.CO},
       adsurl = {https://ui.adsabs.harvard.edu/abs/2025PhRvD.112b3002S}
}

@ARTICLE{2025PhRvD.112b3546L,
       author = {{Lester}, Earl and {Bolejko}, Krzysztof},
        title = "{Constraining decaying dark matter models with gravitational lensing and cosmic voids}",
      journal = {\prd},
         year = 2025,
        month = jul,
       volume = {112},
       number = {2},
          eid = {023546},
        pages = {023546},
          doi = {10.1103/5kk7-7f4n},
archivePrefix = {arXiv},
       eprint = {2507.08275},
 primaryClass = {astro-ph.CO},
       adsurl = {https://ui.adsabs.harvard.edu/abs/2025PhRvD.112b3546L}
}

@ARTICLE{2020MNRAS.494.1956M,
       author = {{Mukherjee}, Suvodip and {Wandelt}, Benjamin D. and {Silk}, Joseph},
        title = "{Probing the theory of gravity with gravitational lensing of gravitational waves and galaxy surveys}",
      journal = {\mnras},
         year = 2020,
        month = may,
       volume = {494},
       number = {2},
        pages = {1956-1970},
          doi = {10.1093/mnras/staa827},
archivePrefix = {arXiv},
       eprint = {1908.08951},
 primaryClass = {astro-ph.CO},
       adsurl = {https://ui.adsabs.harvard.edu/abs/2020MNRAS.494.1956M}
}

@ARTICLE{2021arXiv210909882E,
       author = {{Evans}, Matthew and {Adhikari}, Rana X and {Afle}, Chaitanya and {Ballmer}, Stefan W. and {Biscoveanu}, Sylvia and {Borhanian}, Ssohrab and {Brown}, Duncan A. and {Chen}, Yanbei and {Eisenstein}, Robert and {Gruson}, Alexandra and {Gupta}, Anuradha and {Hall}, Evan D. and {Huxford}, Rachael and {Kamai}, Brittany and {Kashyap}, Rahul and {Kissel}, Jeff S. and {Kuns}, Kevin and {Landry}, Philippe and {Lenon}, Amber and {Lovelace}, Geoffrey and {McCuller}, Lee and {Ng}, Ken K.~Y. and {Nitz}, Alexander H. and {Read}, Jocelyn and {Sathyaprakash}, B.~S. and {Shoemaker}, David H. and {Slagmolen}, Bram J.~J. and {Smith}, Joshua R. and {Srivastava}, Varun and {Sun}, Ling and {Vitale}, Salvatore and {Weiss}, Rainer},
        title = "{A Horizon Study for Cosmic Explorer: Science, Observatories, and Community}",
      journal = {arXiv e-prints},
         year = 2021,
        month = sep,
          eid = {arXiv:2109.09882},
        pages = {arXiv:2109.09882},
          doi = {10.48550/arXiv.2109.09882},
archivePrefix = {arXiv},
       eprint = {2109.09882},
 primaryClass = {astro-ph.IM},
       adsurl = {https://ui.adsabs.harvard.edu/abs/2021arXiv210909882E}
}

@ARTICLE{2025arXiv250312263A,
       author = {{Abac}, Adrian and {Abramo}, Raul and {Albanesi}, Simone and {Albertini}, Angelica and {Agapito}, Alessandro and {Agathos}, Michalis and {Albertus}, Conrado and {Andersson}, Nils and {Andrade}, Tomas and {Andreoni}, Igor and {Angeloni}, Federico and {Antonelli}, Marco and {Antoniadis}, John and {Antonini}, Fabio and {Arca Sedda}, Manuel and {Artale}, M. Celeste and {Ascenzi}, Stefano and {Auclair}, Pierre and {Bachetti}, Matteo and {Badger}, Charles and {Banerjee}, Biswajit and {Barba-Gonzalez}, David and {Barta}, Daniel and {Bartolo}, Nicola and {Bauswein}, Andreas and {Begnoni}, Andrea and {Beirnaert}, Freija and {Bejger}, Michal and {Belgacem}, Enis and {Bellomo}, Nicola and {Bernard}, Laura and {Grazia Bernardini}, Maria and {Bernuzzi}, Sebastiano and {Berry}, Christopher P.~L. and {Berti}, Emanuele and {Bertone}, Gianfranco and {Bettoni}, Dario and {Bezares}, Miguel and {Bhagwat}, Swetha and {Bisero}, Sofia and {Bizouard}, Marie Anne and {Blanco-Pillado}, Jose J. and {Blasi}, Simone and {Bonino}, Alice and {Borghese}, Alice and {Borhanian}, Ssohrab and {Bortolas}, Elisa and {Botticella}, Maria Teresa and {Branchesi}, Marica and {Breschi}, Matteo and {Brito}, Richard and {Brocato}, Enzo and {Broekgaarden}, Floor S. and {Bulik}, Tomasz and {Buonanno}, Alessandra and {Burgio}, Fiorella and {Burrows}, Adam and {Calcagni}, Gianluca and {Canevarolo}, Sofia and {Cappellaro}, Enrico and {Capurri}, Giulia and {Carbone}, Carmelita and {Casadio}, Roberto and {Cayuso}, Ramiro and {Cerda-Duran}, Pablo and {Char}, Prasanta and {Chaty}, Sylvain and {Chiarusi}, Tommaso and {Chruslinska}, Martyna and {Cireddu}, Francesco and {Cole}, Philippa and {Colombo}, Alberto and {Colpi}, Monica and {Compere}, Geoffrey and {Contaldi}, Carlo and {Corman}, Maxence and {Crescimbeni}, Francesco and {Cristallo}, Sergio and {Cuoco}, Elena and {Cusin}, Giulia and {Dal Canton}, Tito and {Dalya}, Gergely and {D'Avanzo}, Paolo and {Davari}, Nazanin and {De Luca}, Valerio and {De Renzis}, Viola and {Della Valle}, Massimo and {Del Pozzo}, Walter and {De Santi}, Federico and {Ludovico De Santis}, Alessio and {Dietrich}, Tim and {Dimastrogiovanni}, Ema and {Domenech}, Guillem and {Doneva}, Daniela and {Drago}, Marco and {Dupletsa}, Ulyana and {Duval}, Hannah and {Dvorkin}, Irina and {Elias-Rosa}, Nancy and {Fairhurst}, Stephen and {Fantina}, Anthea F. and {Fasiello}, Matteo and {Fays}, Maxime and {Fender}, Rob and {Fischer}, Tobias and {Foucart}, Francois and {Fragos}, Tassos and {Foffa}, Stefano and {Franciolini}, Gabriele and {Fumagalli}, Jacopo and {Gair}, Jonathan and {Gamba}, Rossella and {Garcia-Bellido}, Juan and {Garcia-Quiros}, Cecilio and {Arpad Gergely}, Laszlo and {Ghirlanda}, Giancarlo and {Ghosh}, Archisman and {Giacomazzo}, Bruno and {Gittins}, Fabian and {Giudice}, Ines Francesca and {Goncharov}, Boris and {Gonzalez}, Alejandra and {Goriely}, Stephane and {Graziani}, Luca and {Greco}, Giuseppe and {Gualtieri}, Leonardo and {Guidi}, Gianluca Maria and {Gupta}, Ish and {Haney}, Maria and {Hannam}, Mark and {Harms}, Jan and {Harutyunyan}, Arus and {Haskell}, Brynmor and {Haungs}, Andreas and {Hazra}, Nandini and {Hemming}, Gary and {Heng}, Ik Siong and {Hinderer}, Tanja and {van der Horst}, Alexander and {Hu}, Qian and {Husa}, Sascha and {Iacovelli}, Francesco and {Illuminati}, Giulia and {Inguglia}, Gianluca and {Izquierdo Villalba}, David and {Janquart}, Justin and {Janssens}, Kamiel and {Jenkins}, Alexander C. and {Jones}, Ian and {Kacskovics}, Balazs and {Klessen}, Ralf S. and {Kokkotas}, Kostas and {Kuan}, Hao-Jui and {Kumar}, Sumit and {Kuroyanagi}, Sachiko and {Laghi}, Danny and {Lamberts}, Astrid and {Lambiase}, Gaetano and {Larrouturou}, Francois and {Leaci}, Paola and {Lenzi}, Michele and {Levan}, Andrew and {Li}, T.~G.~F. and {Li}, Yufeng and {Liang}, Dicong and {Limongi}, Marco and {Liu}, Boyuan and {Llanes-Estrada}, Felipe J. and {Loffredo}, Eleonora and {Long}, Oliver and {Lope-Oter}, Eva and {Lukes-Gerakopoulos}, Georgios and {Maggio}, Elisa and {Maggiore}, Michele and {Mancarella}, Michele and {Mapelli}, Michela and {Marchant}, Pablo and {Margiotta}, Annarita and {Mariotti}, Alberto and {Marriott-Best}, Alisha and {Marsat}, Sylvain and {Martinez-Pinedo}, Gabriel and {Maselli}, Andrea and {Mastrogiovanni}, Simone and {Matos}, Isabela and {Melandri}, Andrea and {Mendes}, Raissa F.~P. and {Mendonca Soares de Souza}, Josiel and {Mentasti}, Giorgio and {Mezcua}, Mar and {Mosta}, Philipp and {Mondal}, Chiranjib and {Moresco}, Michele and {Mukherjee}, Tista and {Muttoni}, Niccolo and {Nagar}, Alessandro and {Narola}, Harsh and {Nava}, Lara and {Navarro Moreno}, Pablo and {Nelemans}, Gijs},
        title = "{The Science of the Einstein Telescope}",
      journal = {arXiv e-prints},
         year = 2025,
        month = mar,
          eid = {arXiv:2503.12263},
        pages = {arXiv:2503.12263},
          doi = {10.48550/arXiv.2503.12263},
archivePrefix = {arXiv},
       eprint = {2503.12263},
 primaryClass = {gr-qc},
       adsurl = {https://ui.adsabs.harvard.edu/abs/2025arXiv250312263A}
}

@ARTICLE{2017arXiv170200786A,
       author = {{Amaro-Seoane}, Pau and {Audley}, Heather and {Babak}, Stanislav and {Baker}, John and {Barausse}, Enrico and {Bender}, Peter and {Berti}, Emanuele and {Binetruy}, Pierre and {Born}, Michael and {Bortoluzzi}, Daniele and {Camp}, Jordan and {Caprini}, Chiara and {Cardoso}, Vitor and {Colpi}, Monica and {Conklin}, John and {Cornish}, Neil and {Cutler}, Curt and {Danzmann}, Karsten and {Dolesi}, Rita and {Ferraioli}, Luigi and {Ferroni}, Valerio and {Fitzsimons}, Ewan and {Gair}, Jonathan and {Gesa Bote}, Lluis and {Giardini}, Domenico and {Gibert}, Ferran and {Grimani}, Catia and {Halloin}, Hubert and {Heinzel}, Gerhard and {Hertog}, Thomas and {Hewitson}, Martin and {Holley-Bockelmann}, Kelly and {Hollington}, Daniel and {Hueller}, Mauro and {Inchauspe}, Henri and {Jetzer}, Philippe and {Karnesis}, Nikos and {Killow}, Christian and {Klein}, Antoine and {Klipstein}, Bill and {Korsakova}, Natalia and {Larson}, Shane L and {Livas}, Jeffrey and {Lloro}, Ivan and {Man}, Nary and {Mance}, Davor and {Martino}, Joseph and {Mateos}, Ignacio and {McKenzie}, Kirk and {McWilliams}, Sean T and {Miller}, Cole and {Mueller}, Guido and {Nardini}, Germano and {Nelemans}, Gijs and {Nofrarias}, Miquel and {Petiteau}, Antoine and {Pivato}, Paolo and {Plagnol}, Eric and {Porter}, Ed and {Reiche}, Jens and {Robertson}, David and {Robertson}, Norna and {Rossi}, Elena and {Russano}, Giuliana and {Schutz}, Bernard and {Sesana}, Alberto and {Shoemaker}, David and {Slutsky}, Jacob and {Sopuerta}, Carlos F. and {Sumner}, Tim and {Tamanini}, Nicola and {Thorpe}, Ira and {Troebs}, Michael and {Vallisneri}, Michele and {Vecchio}, Alberto and {Vetrugno}, Daniele and {Vitale}, Stefano and {Volonteri}, Marta and {Wanner}, Gudrun and {Ward}, Harry and {Wass}, Peter and {Weber}, William and {Ziemer}, John and {Zweifel}, Peter},
        title = "{Laser Interferometer Space Antenna}",
      journal = {arXiv e-prints},
         year = 2017,
        month = feb,
          eid = {arXiv:1702.00786},
        pages = {arXiv:1702.00786},
          doi = {10.48550/arXiv.1702.00786},
archivePrefix = {arXiv},
       eprint = {1702.00786},
 primaryClass = {astro-ph.IM},
       adsurl = {https://ui.adsabs.harvard.edu/abs/2017arXiv170200786A}
}

@ARTICLE{2016CQGra..33c5010L,
       author = {{Luo}, Jun and {Chen}, Li-Sheng and {Duan}, Hui-Zong and {Gong}, Yun-Gui and {Hu}, Shoucun and {Ji}, Jianghui and {Liu}, Qi and {Mei}, Jianwei and {Milyukov}, Vadim and {Sazhin}, Mikhail and {Shao}, Cheng-Gang and {Toth}, Viktor T. and {Tu}, Hai-Bo and {Wang}, Yamin and {Wang}, Yan and {Yeh}, Hsien-Chi and {Zhan}, Ming-Sheng and {Zhang}, Yonghe and {Zharov}, Vladimir and {Zhou}, Ze-Bing},
        title = "{TianQin: a space-borne gravitational wave detector}",
      journal = {Classical and Quantum Gravity},
         year = 2016,
        month = feb,
       volume = {33},
       number = {3},
          eid = {035010},
        pages = {035010},
          doi = {10.1088/0264-9381/33/3/035010},
archivePrefix = {arXiv},
       eprint = {1512.02076},
 primaryClass = {astro-ph.IM},
       adsurl = {https://ui.adsabs.harvard.edu/abs/2016CQGra..33c5010L}
}

@ARTICLE{2020IJMPA..3550075R,
       author = {{Ruan}, Wen-Hong and {Guo}, Zong-Kuan and {Cai}, Rong-Gen and {Zhang}, Yuan-Zhong},
        title = "{Taiji program: Gravitational-wave sources}",
      journal = {International Journal of Modern Physics A},
         year = 2020,
        month = jun,
       volume = {35},
       number = {17},
          eid = {2050075},
        pages = {2050075},
          doi = {10.1142/S0217751X2050075X},
archivePrefix = {arXiv},
       eprint = {1807.09495},
 primaryClass = {gr-qc},
       adsurl = {https://ui.adsabs.harvard.edu/abs/2020IJMPA..3550075R}
}

@ARTICLE{2010PhRvL.105y1101S,
       author = {{Sereno}, M. and {Sesana}, A. and {Bleuler}, A. and {Jetzer}, Ph. and {Volonteri}, M. and {Begelman}, M.~C.},
        title = "{Strong Lensing of Gravitational Waves as Seen by LISA}",
      journal = {\prl},
         year = 2010,
        month = dec,
       volume = {105},
       number = {25},
          eid = {251101},
        pages = {251101},
          doi = {10.1103/PhysRevLett.105.251101},
archivePrefix = {arXiv},
       eprint = {1011.5238},
 primaryClass = {astro-ph.CO},
       adsurl = {https://ui.adsabs.harvard.edu/abs/2010PhRvL.105y1101S}
}

@ARTICLE{2020A&A...641A...6P,
       author = {{Planck Collaboration} and {Aghanim}, N. and {Akrami}, Y. and {Ashdown}, M. and {Aumont}, J. and {Baccigalupi}, C. and {Ballardini}, M. and {Banday}, A.~J. and {Barreiro}, R.~B. and {Bartolo}, N. and {Basak}, S. and {Battye}, R. and {Benabed}, K. and {Bernard}, J.-P. and {Bersanelli}, M. and {Bielewicz}, P. and {Bock}, J.~J. and {Bond}, J.~R. and {Borrill}, J. and {Bouchet}, F.~R. and {Boulanger}, F. and {Bucher}, M. and {Burigana}, C. and {Butler}, R.~C. and {Calabrese}, E. and {Cardoso}, J.-F. and {Carron}, J. and {Challinor}, A. and {Chiang}, H.~C. and {Chluba}, J. and {Colombo}, L.~P.~L. and {Combet}, C. and {Contreras}, D. and {Crill}, B.~P. and {Cuttaia}, F. and {de Bernardis}, P. and {de Zotti}, G. and {Delabrouille}, J. and {Delouis}, J.-M. and {Di Valentino}, E. and {Diego}, J.~M. and {Dor{\'e}}, O. and {Douspis}, M. and {Ducout}, A. and {Dupac}, X. and {Dusini}, S. and {Efstathiou}, G. and {Elsner}, F. and {En{\ss}lin}, T.~A. and {Eriksen}, H.~K. and {Fantaye}, Y. and {Farhang}, M. and {Fergusson}, J. and {Fernandez-Cobos}, R. and {Finelli}, F. and {Forastieri}, F. and {Frailis}, M. and {Fraisse}, A.~A. and {Franceschi}, E. and {Frolov}, A. and {Galeotta}, S. and {Galli}, S. and {Ganga}, K. and {G{\'e}nova-Santos}, R.~T. and {Gerbino}, M. and {Ghosh}, T. and {Gonz{\'a}lez-Nuevo}, J. and {G{\'o}rski}, K.~M. and {Gratton}, S. and {Gruppuso}, A. and {Gudmundsson}, J.~E. and {Hamann}, J. and {Handley}, W. and {Hansen}, F.~K. and {Herranz}, D. and {Hildebrandt}, S.~R. and {Hivon}, E. and {Huang}, Z. and {Jaffe}, A.~H. and {Jones}, W.~C. and {Karakci}, A. and {Keih{\"a}nen}, E. and {Keskitalo}, R. and {Kiiveri}, K. and {Kim}, J. and {Kisner}, T.~S. and {Knox}, L. and {Krachmalnicoff}, N. and {Kunz}, M. and {Kurki-Suonio}, H. and {Lagache}, G. and {Lamarre}, J.-M. and {Lasenby}, A. and {Lattanzi}, M. and {Lawrence}, C.~R. and {Le Jeune}, M. and {Lemos}, P. and {Lesgourgues}, J. and {Levrier}, F. and {Lewis}, A. and {Liguori}, M. and {Lilje}, P.~B. and {Lilley}, M. and {Lindholm}, V. and {L{\'o}pez-Caniego}, M. and {Lubin}, P.~M. and {Ma}, Y.-Z. and {Mac{\'\i}as-P{\'e}rez}, J.~F. and {Maggio}, G. and {Maino}, D. and {Mandolesi}, N. and {Mangilli}, A. and {Marcos-Caballero}, A. and {Maris}, M. and {Martin}, P.~G. and {Martinelli}, M. and {Mart{\'\i}nez-Gonz{\'a}lez}, E. and {Matarrese}, S. and {Mauri}, N. and {McEwen}, J.~D. and {Meinhold}, P.~R. and {Melchiorri}, A. and {Mennella}, A. and {Migliaccio}, M. and {Millea}, M. and {Mitra}, S. and {Miville-Desch{\^e}nes}, M.-A. and {Molinari}, D. and {Montier}, L. and {Morgante}, G. and {Moss}, A. and {Natoli}, P. and {N{\o}rgaard-Nielsen}, H.~U. and {Pagano}, L. and {Paoletti}, D. and {Partridge}, B. and {Patanchon}, G. and {Peiris}, H.~V. and {Perrotta}, F. and {Pettorino}, V. and {Piacentini}, F. and {Polastri}, L. and {Polenta}, G. and {Puget}, J.-L. and {Rachen}, J.~P. and {Reinecke}, M. and {Remazeilles}, M. and {Renzi}, A. and {Rocha}, G. and {Rosset}, C. and {Roudier}, G. and {Rubi{\~n}o-Mart{\'\i}n}, J.~A. and {Ruiz-Granados}, B. and {Salvati}, L. and {Sandri}, M. and {Savelainen}, M. and {Scott}, D. and {Shellard}, E.~P.~S. and {Sirignano}, C. and {Sirri}, G. and {Spencer}, L.~D. and {Sunyaev}, R. and {Suur-Uski}, A.-S. and {Tauber}, J.~A. and {Tavagnacco}, D. and {Tenti}, M. and {Toffolatti}, L. and {Tomasi}, M. and {Trombetti}, T. and {Valenziano}, L. and {Valiviita}, J. and {Van Tent}, B. and {Vibert}, L. and {Vielva}, P. and {Villa}, F. and {Vittorio}, N. and {Wandelt}, B.~D. and {Wehus}, I.~K. and {White}, M. and {White}, S.~D.~M. and {Zacchei}, A. and {Zonca}, A.},
        title = "{Planck 2018 results. VI. Cosmological parameters}",
      journal = {\aap},
         year = 2020,
        month = sep,
       volume = {641},
          eid = {A6},
        pages = {A6},
          doi = {10.1051/0004-6361/201833910},
archivePrefix = {arXiv},
       eprint = {1807.06209},
 primaryClass = {astro-ph.CO},
       adsurl = {https://ui.adsabs.harvard.edu/abs/2020A&A...641A...6P}
}

@misc{Abac_2025_GWTC4,
    title = {{GWTC-4.0}: Population Properties of Merging Compact Binaries},
    author = {{The LIGO Scientific Collaboration} and {The Virgo Collaboration} and {The KAGRA Collaboration} and Abac, A. G. and others},
    year = {2025},
    eprint = {2508.18083},
    archivePrefix = {arXiv},
    primaryClass = {astro-ph.HE},
    note = {arXiv:2508.18083}
}

@ARTICLE{2023MNRAS.518.5298B,
       author = {{Biscoveanu}, Sylvia and {Landry}, Philippe and {Vitale}, Salvatore},
        title = "{Population properties and multimessenger prospects of neutron star-black hole mergers following GWTC-3}",
      journal = {\mnras},
         year = 2023,
        month = feb,
       volume = {518},
       number = {4},
        pages = {5298-5312},
          doi = {10.1093/mnras/stac3052},
archivePrefix = {arXiv},
       eprint = {2207.01568},
 primaryClass = {astro-ph.HE},
       adsurl = {https://ui.adsabs.harvard.edu/abs/2023MNRAS.518.5298B}
}

@ARTICLE{2019ApJ...871..168D,
       author = {{Diemer}, Benedikt and {Joyce}, Michael},
        title = "{An Accurate Physical Model for Halo Concentrations}",
      journal = {\apj},
         year = 2019,
        month = feb,
       volume = {871},
       number = {2},
          eid = {168},
        pages = {168},
          doi = {10.3847/1538-4357/aafad6},
archivePrefix = {arXiv},
       eprint = {1809.07326},
 primaryClass = {astro-ph.CO},
       adsurl = {https://ui.adsabs.harvard.edu/abs/2019ApJ...871..168D}
}

@ARTICLE{2021PASP..133g4504O,
       author = {{Oguri}, Masamune},
        title = "{Fast Calculation of Gravitational Lensing Properties of Elliptical Navarro-Frenk-White and Hernquist Density Profiles}",
      journal = {\pasp},
         year = 2021,
        month = jul,
       volume = {133},
       number = {1025},
          eid = {074504},
        pages = {074504},
          doi = {10.1088/1538-3873/ac12db},
archivePrefix = {arXiv},
       eprint = {2106.11464},
 primaryClass = {astro-ph.IM},
       adsurl = {https://ui.adsabs.harvard.edu/abs/2021PASP..133g4504O}
}

@ARTICLE{Piorkowska2021,
       author = {{Pi{\'o}rkowska-Kurpas}, Aleksandra and {Hou}, Shaoqi and {Biesiada}, Marek and {Ding}, Xuheng and {Cao}, Shuo and {Fan}, Xilong and {Kawamura}, Seiji and {Zhu}, Zong-Hong},
        title = "{Inspiraling Double Compact Object Detection and Lensing Rate: Forecast for DECIGO and B-DECIGO}",
      journal = {\apj},
         year = 2021,
        month = feb,
       volume = {908},
       number = {2},
          eid = {196},
        pages = {196},
          doi = {10.3847/1538-4357/abd482},
archivePrefix = {arXiv},
       eprint = {2005.08727},
 primaryClass = {astro-ph.HE},
       adsurl = {https://ui.adsabs.harvard.edu/abs/2021ApJ...908..196P}
}

@ARTICLE{2021PhRvD.103h3011M,
       author = {{Marsat}, Sylvain and {Baker}, John G. and {Canton}, Tito Dal},
        title = "{Exploring the Bayesian parameter estimation of binary black holes with LISA}",
      journal = {\prd},
         year = 2021,
        month = apr,
       volume = {103},
       number = {8},
          eid = {083011},
        pages = {083011},
          doi = {10.1103/PhysRevD.103.083011},
archivePrefix = {arXiv},
       eprint = {2003.00357},
 primaryClass = {gr-qc},
       adsurl = {https://ui.adsabs.harvard.edu/abs/2021PhRvD.103h3011M}
}

@ARTICLE{2019ApJ...880...50Y,
       author = {{Yang}, Tao and {Hu}, Bin and {Cai}, Rong-Gen and {Wang}, Bin},
        title = "{New Probe of Gravity: Strongly Lensed Gravitational-wave Multimessenger Approach}",
      journal = {\apj},
         year = 2019,
        month = jul,
       volume = {880},
       number = {1},
          eid = {50},
        pages = {50},
          doi = {10.3847/1538-4357/ab271e},
archivePrefix = {arXiv},
       eprint = {1810.00164},
 primaryClass = {astro-ph.CO},
       adsurl = {https://ui.adsabs.harvard.edu/abs/2019ApJ...880...50Y}
}

@ARTICLE{2020MNRAS.497L..56Y,
       author = {{Yang}, Tao and {Birrer}, Simon and {Hu}, Bin},
        title = "{The first simultaneous measurement of Hubble constant and post-Newtonian parameter from time-delay strong lensing}",
      journal = {\mnras},
         year = 2020,
        month = sep,
       volume = {497},
       number = {1},
        pages = {L56-L61},
          doi = {10.1093/mnrasl/slaa107},
archivePrefix = {arXiv},
       eprint = {2003.03277},
 primaryClass = {astro-ph.CO},
       adsurl = {https://ui.adsabs.harvard.edu/abs/2020MNRAS.497L..56Y}
}

@ARTICLE{2025arXiv250509507Y,
       author = {{Ying}, Xinguang and {Yang}, Tao},
        title = "{Precision Joint Constraints on Cosmology and Gravity Using Strongly Lensed Gravitational Wave Populations}",
      journal = {arXiv e-prints},
         year = 2025,
        month = may,
          eid = {arXiv:2505.09507},
        pages = {arXiv:2505.09507},
          doi = {10.48550/arXiv.2505.09507},
archivePrefix = {arXiv},
       eprint = {2505.09507},
 primaryClass = {gr-qc},
       adsurl = {https://ui.adsabs.harvard.edu/abs/2025arXiv250509507Y}
}

@dataset{lensedgw_2026_20540878,
  author       = {LensedGW},
  title        = {LensedGW/GW-LMC-Space: GW-LMC-Space},
  month        = jun,
  year         = 2026,
  publisher    = {Zenodo},
  version      = {v2.0},
  doi          = {10.5281/zenodo.20540878},
  url          = {https://doi.org/10.5281/zenodo.20540878},
}

@article{bilby_paper,
    author = "Ashton, Gregory and others",
    title = "{BILBY: A user-friendly Bayesian inference library for gravitational-wave astronomy}",
    eprint = "1811.02042",
    archivePrefix = "arXiv",
    primaryClass = "astro-ph.IM",
    doi = "10.3847/1538-4365/ab06fc",
    journal = "Astrophys. J. Suppl.",
    volume = "241",
    number = "2",
    pages = "27",
    year = "2019"
}

@software{colm_talbot_2025_17533961,
  author       = {Colm Talbot and
                  Gregory Ashton and
                  Moritz Hübner and
                  Matt Pitkin and
                  plasky and
                  Michael J. Williams and
                  asb5468 and
                  Aditya Vijaykumar and
                  Rory Smith and
                  SMorisaki and
                  John Veitch and
                  Nikhil Sarin and
                  Duncan Macleod and
                  Daniel Williams and
                  JasperMartins and
                  MarcArene and
                  C P L Berry and
                  Vivien Raymond and
                  Ceciliogq and
                  Ivan Markin and
                  David Keitel and
                  AlexandreGoettel and
                  Lorenzo Pompili and
                  Mick Wright and
                  oliviawilk and
                  noahewolfe and
                  jacobgolomb and
                  Shichao Wu and
                  Rhiannon Udall and
                  Michael Pürrer},
  title        = {bilby-dev/bilby: v.2.7.1},
  month        = nov,
  year         = 2025,
  publisher    = {Zenodo},
  version      = {v2.7.1},
  doi          = {10.5281/zenodo.17533961},
  url          = {https://doi.org/10.5281/zenodo.17533961},
  swhid        = {swh:1:dir:3840032346ac7f56582e6c74cc48dccc33747418
                   ;origin=https://doi.org/10.5281/zenodo.14025463;vi
                   sit=swh:1:snp:b4b0318c71303094707fd61e3305dd2addf1
                   a743;anchor=swh:1:rel:7b67a76d1e7e224899d337492665
                   5fa8e0e7074c;path=bilby-dev-bilby-710e180
                  },
}

@ARTICLE{2022JCAP...07..022B,
       author = {{Bulashenko}, Oleg and {Ubach}, Helena},
        title = "{Lensing of gravitational waves: universal signatures in the beating pattern}",
      journal = {\jcap},
         year = 2022,
        month = jul,
       volume = {2022},
       number = {7},
          eid = {022},
        pages = {022},
          doi = {10.1088/1475-7516/2022/07/022},
archivePrefix = {arXiv},
       eprint = {2112.10773},
 primaryClass = {gr-qc},
       adsurl = {https://ui.adsabs.harvard.edu/abs/2022JCAP...07..022B}
}

\end{document}